\documentclass[12pt,letterpaper]{article}
\usepackage{epsfig,rotating,setspace,latexsym,amsmath,epsf,amssymb,bm,theorem}
\usepackage{cite,appendix}

\usepackage{amsfonts}

\title{An Outer Bound for the Vector Gaussian CEO Problem\thanks{This work was supported by NSF Grants CNS 09-64632, CCF
09-64645, CCF 10-18185, and CNS 11-47811.}}
\author{Ersen Ekrem \qquad Sennur Ulukus
\\
\normalsize Department of Electrical and Computer Engineering\\
\normalsize University of Maryland, College Park, MD 20742 \\
\normalsize {\it ersen@umd.edu} \qquad {\it ulukus@umd.edu}}

\newcommand{\mmse}{{\rm mmse}}

\newcommand{\bblambda}{\bm \Lambda}

\newcommand{\bbsigma}{\bm \Sigma}

\newcommand{\bbi}{{\mathbf{I}}}
\newcommand{\bzero}{{\mathbf{0}}}
\newcommand{\bbv}{{\mathbf{V}}}

\newcommand{\bbh}{{\mathbf{H}}}

\newcommand{\bbm}{{\mathbf{M}}}

\newcommand{\bbk}{{\mathbf{K}}}

\newcommand{\bbn}{{\mathbf{N}}}

\newcommand{\bba}{{\mathbf{A}}}
\newcommand{\bbd}{{\mathbf{D}}}

\newcommand{\bbt}{{\mathbf{T}}}

\newcommand{\bbc}{{\mathbf{C}}}

\newcommand{\bbs}{{\mathbf{S}}}

\newcommand{\bbj}{{\mathbf{J}}}
\newcommand{\bbu}{{\mathbf{U}}}
\newcommand{\bx}{{\mathbf{x}}}
\newcommand{\bbx}{{\mathbf{X}}}
\newcommand{\by}{{\mathbf{y}}}
\newcommand{\bby}{{\mathbf{Y}}}

\newtheorem{Theo}{Theorem}

\newtheorem{Lem}{Lemma}
\newtheorem{Cor}{Corollary}

\begin{document}

%\IEEEoverridecommandlockouts

\maketitle

\begin{abstract}
We study the vector Gaussian CEO problem, where there are an
arbitrary number of agents each having a noisy observation of a
vector Gaussian source. The goal of the agents is to describe the
source to a central unit, which wants to reconstruct the source
within a given distortion. The rate-distortion region of the
vector Gaussian CEO problem is unknown in general. Here, we
provide an outer bound for the rate-distortion region of the
vector Gaussian CEO problem. We obtain our outer bound by
evaluating an outer bound for the multi-terminal source coding
problem by means of a technique relying on the de Bruijn identity
and the properties of the Fisher information. Next, we show that
our outer bound strictly improves upon the existing outer bounds
for all system parameters. We show this strict improvement by
providing a specific example, and showing that there exists a gap
between our outer bound and the existing outer bounds. Although
our outer bound improves upon the existing outer bounds, we show
that our outer bound does not provide the exact rate-distortion
region in general. To this end, we provide an example and show
that the rate-distortion region is strictly contained in our outer
bound for this example.
\end{abstract}

\newpage

\section{Introduction}

\begin{figure}[t]
\centering
\includegraphics[width=6in]{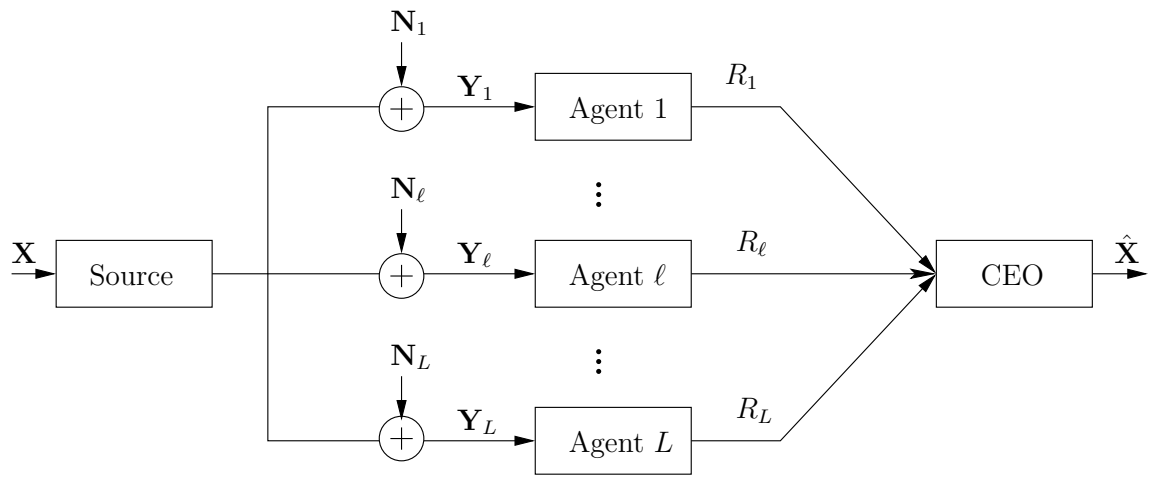}
\caption{The vector Gaussian CEO problem.}
\label{Vector_CEO_Model}
\end{figure}

We study the vector Gaussian CEO problem, where there is a vector
Gaussian source which is observed through some noisy channels by
an arbitrary number of agents. The agents process their
observations independently and communicate them to a central unit
(the so-called CEO unit) through orthogonal and rate-limited links
(see Figure~\ref{Vector_CEO_Model}). The goal of the agents is to
describe their observations to the central unit in a way that the
central unit can reconstruct the source within a given distortion.
The fundamental trade-off between the rate spent by the agents to
describe the source and the distortion attained by the central
unit is characterized by the rate-distortion region, which is
unknown in general.

The CEO problem is introduced in~\cite{CEO_Introduction}, where
the authors consider a discrete memoryless setting where the
source and the observations of the agents all come from some
discrete alphabet. In the setting of~\cite{CEO_Introduction}, the
central unit is interested in estimating the source with the
minimum expected error frequency which corresponds to the Hamming
distance between the source sequence and the central unit's
estimation of the source sequence. In~\cite{CEO_Introduction}, the
authors consider the decay rate of the error frequency with
respect to the rate expenditure of the agents, and obtain the best
possible decay rate when the number of agents goes to infinity.

The scalar Gaussian CEO problem is studied
in~\cite{Quadratic_Gaussian_CEO}, where there is a scalar Gaussian
source which is observed through some linear Gaussian channels by
the agents. The agents describe their observations to the central
unit in a way that the central unit can reconstruct the source
within a certain minimum mean square error (MMSE).
In~\cite{Quadratic_Gaussian_CEO}, the decay rate of the MMSE with
respect to the rate expenditure of the agents is considered and
shown to be inversely proportional with the rate expenditure of
the agents, when the number of agents goes to infinity. The scalar
Gaussian CEO problem is further studied
in~\cite{Oohama_CEO,Prabhakaran_CEO}, where instead of the decay
rate of the achievable MMSE, the focus was on the entire
rate-distortion region. In~\cite{Oohama_CEO,Prabhakaran_CEO}, the
entire rate-distortion region for the scalar Gaussian problem is
established. The achievability is shown by using the Berger-Tung
inner bound~\cite{thesis_tung}, and the converse is established by
using the entropy-power inequality. Recently, an alternative proof
for the sum-rate of the scalar Gaussian CEO problem is established
in~\cite{Chen_alternative} without invoking the entropy-power
inequality.

As pointed out by several
works~\cite{Shamai_MIMO,MIMO_BC_Secrecy}, although entropy-power
inequality is a key tool in providing converse proofs for scalar
Gaussian problems, it might be restrictive for vector Gaussian
problems. For the vector Gaussian CEO problem, this observation is
noticed in~\cite{Vector_CEO_sum_rate}, where the authors provide a
lower bound for the sum-rate of the vector Gaussian CEO problem by
using the entropy-power inequality. This lower bound is shown to
be tight under certain conditions, although it is not tight in
general. Recently,~\cite{Chen_Wang_Vector_CEO} provided an outer
bound for the rate-distortion region of the vector Gaussian CEO
problem when there are only two agents. They obtain their outer
bound by using an extremal inequality, which can be viewed as a
generalization of the extremal inequality provided
in~\cite{Liu_Extremal_Inequality}.

In this paper, we consider the vector Gaussian CEO problem for an
arbitrary number of agents and provide an outer bound for its
rate-distortion region. We first consider the outer bound provided
in~\cite{Wagner_Outer_Bound} for the multi-terminal source coding
problem, and evaluate it for the vector Gaussian CEO problem at
hand. In the evaluation of the outer bound
in~\cite{Wagner_Outer_Bound}, we use the de Bruijn
identity~\cite{Palomar_Gradient}, a connection between the
differential entropy and the Fisher information, along with the
properties of the MMSE and the Fisher information. This evaluation
technique which relies on the de Bruijn identity is useful in the
sense that it is able to alleviate some shortcomings of the
entropy-power inequality in vector Gaussian
problems~\cite{MIMO_BC_Secrecy,Ekrem_Ulukus_Alternative}.

Next, we compare our outer bound with the best known outer bound
for the rate-distortion region of the vector Gaussian CEO problem
given in~\cite{Chen_Wang_Vector_CEO}. We show that the outer bound
in~\cite{Chen_Wang_Vector_CEO} contains our outer bound in
general, for all system parameters. We then provide a specific
example where the outer bound in~\cite{Chen_Wang_Vector_CEO}
strictly contains our outer bound. In other words, our outer bound
brings a strict improvement over the outer bound
in~\cite{Chen_Wang_Vector_CEO}. However, in spite of this strict
improvement, our outer bound falls short of providing the exact
rate-distortion region of the vector Gaussian CEO problem in
general. We establish this fact by considering the parallel
Gaussian model, for which we obtain the entire rate-distortion
region explicitly and show that our outer bound strictly includes
this rate-distortion region. In other words, for the parallel
Gaussian model, our outer bound is not equal to the
rate-distortion region, which shows that our outer bound is not
tight in general.

\section{Problem Statement and the Main Result}
\label{sec:aligned_model}

In the CEO problem, there are $L$ sensors, each of which getting a
noisy observation of a source. The goal of the sensors is to
describe their observations to the CEO unit such that the CEO unit
can reconstruct the source within a given distortion. In the
vector Gaussian CEO problem, there is an i.i.d. vector Gaussian
source $\{\bbx_i\}_{i=1}^n$ with zero-mean and covariance
$\bbk_X$. Each sensor gets a noisy version of this Gaussian source
\begin{align}
\bby_{\ell,i}=\bbx_{i}+\bbn_{\ell,i},\quad \ell=1,\ldots,L
\label{aligned_observations}
\end{align}
where $\{\bbn_{\ell,i}\}_{i=1}^n$ is an i.i.d. sequence of
Gaussian random vectors with zero-mean and covariance
$\bbsigma_{\ell}$. Moreover, noise among the sensors are
independent, i.e., $\bbn_{1,i},\ldots,\bbn_{L,i}$ are independent
$\forall i=1,\ldots,n$. In the vector Gaussian CEO problem, the
distortion of the reconstructed vector is measured by its mean
square error matrix
\begin{align}
\frac{1}{n}\sum_{i=1}^n
E\left[\big(\bbx_i-\hat{\bbx}_i\big)\big(\bbx_i-\hat{\bbx}_i\big)^\top\right]
\label{distortion_measure}
\end{align}
where $\hat{\bbx}^n$ denotes the reconstructed vector.

An $(n,R_1,\ldots,R_L)$ code for the CEO problem consists of an
encoding function at each sensor $f_\ell^n:\mathbb{R}^{M\times
n}\rightarrow \mathcal{B}_\ell^n=\{1,\ldots,2^{nR_\ell}\}$, i.e.,
$B_\ell^n=f_\ell^n(\bby_\ell^n)$ where
$B_\ell^n\in\mathcal{B}_\ell^n,~\ell=1,\ldots,L$, and a decoding
function at the CEO unit
$g^n:\mathcal{B}_{1}^n\times\ldots\times\mathcal{B}_L^n\rightarrow
\mathbb{R}^{M\times n}$, i.e.,
$\hat{\bbx}^n=g^n(B_1^n,\ldots,B_L^n)$, where $M$ denotes the size
of the vector Gaussian source $\bbx$.

We note that since the mean square error is minimized by the MMSE
estimator, which is the conditional mean, without loss of
generality, the decoding function $g^{n}$ can be chosen as the
MMSE estimator. Consequently, we have
\begin{align}
\hat{\bbx}_i=E\left[\bbx_i|B_1^n,\ldots,B_L^n\right]
\end{align}
using which in (\ref{distortion_measure}), we get
\begin{align}
\frac{1}{n}\sum_{i=1}^n
E\left[\big(\bbx_i-\hat{\bbx}_i\big)\big(\bbx_i-\hat{\bbx}_i\big)^\top\right]&=\frac{1}{n}\sum_{i=1}^n\mmse(\bbx_i|B_1^n,\ldots,B_L^n)
\label{new_distortion_measure}
\end{align}
In view of (\ref{new_distortion_measure}), a rate tuple
$(R_1,\ldots,R_L)$ is said to achieve the distortion $\bbd$ if
there exists an $(n,R_1,\ldots,R_L)$ code such that
\begin{align}
\lim_{n\rightarrow
\infty}\frac{1}{n}\sum_{i=1}^n\mmse(\bbx_i|B_1^n,\ldots,B_L^n)
\preceq \bbd
\end{align}
where $\bbd$ is a strictly positive definite matrix. Throughout
the paper, we assume that the distortion matrix $\bbd$ satisfies
\begin{align}
\left(\bbk_X^{-1}+\sum_{\ell=1}^L \bbsigma_\ell^{-1}\right)^{-1}
\preceq \bbd \preceq \bbk_X \label{limits_of_distortion}
\end{align}
where the lower bound on the distortion constraint $\bbd$
corresponds to the MMSE matrix obtained when the CEO unit has
direct access to the observations of the agents
$\{\bby_\ell\}_{\ell=1}^L$. The derivation of this lower bound is
provided in Appendix~\ref{appendix_distortion_limits}, where we
also provide insight on the upper bound in
(\ref{limits_of_distortion}). In
Appendix~\ref{appendix_distortion_limits}, we also show that
imposing the lower bound on $\bbd$ in
(\ref{limits_of_distortion}), i.e., imposing
$\left(\bbk_X^{-1}+\sum_{\ell=1}^L \bbsigma_\ell^{-1}\right)^{-1}
\preceq \bbd$, does not incur any loss of generality, while
imposing the upper bound on $\bbd$ in
(\ref{limits_of_distortion}), i.e., imposing $\bbd\preceq \bbk_X$,
might incur some loss of generality.

The rate-distortion region $\mathcal{R}(\bbd)$ of the vector
Gaussian CEO problem is defined as the closure of all rate tuples
$(R_1,\ldots,R_L)$ that can achieve the distortion $\bbd$.

The main result of this paper is the following outer bound on the
rate-distortion region $\mathcal{R}(\bbd)$ of the vector Gaussian
CEO problem stated in the following theorem.
\begin{Theo}
\label{theorem_outer} The rate-distortion region of the Gaussian
CEO problem $\mathcal{R}(\bbd)$ is contained in the region
$\mathcal{R}^{o}(\bbd)$ which is given by the union of rate tuples
$(R_1,\ldots,R_L)$ satisfying
\begin{align}
\sum_{\ell\in\mathcal{A}}R_\ell\geq \frac{1}{2} \log^+\frac{\left|
\left(\bbk_X^{-1}+\sum_{\ell\in\mathcal{A}^c}\bbsigma_\ell^{-1}-\sum_{\ell\in\mathcal{A}^c}
\bbsigma_\ell^{-1} \bbd_\ell\bbsigma_\ell^{-1}\right)^{-1}
\right|}{|\bbd|}+\sum_{\ell\in\mathcal{A}}\frac{1}{2}
\log\frac{|\bbsigma_\ell|}{|\bbd_\ell|} \label{rate_bound_outer}
\end{align}
for all $\mathcal{A}\subseteq\{1,\ldots,L\}$, where the union is
over all positive semi-definite matrices
$\{\bbd_\ell\}_{\ell=1}^L$ satisfying the following constraints
\begin{align}
\left(
\bbk_X^{-1}+\sum_{\ell=1}^L\bbsigma_\ell^{-1}-\sum_{\ell=1}^{L}
\bbsigma_\ell^{-1} \bbd_\ell\bbsigma_\ell^{-1}\right)^{-1} &\preceq \bbd \label{feasible_set_outer_I}\\
\bzero \preceq \bbd_\ell &\preceq \bbsigma_\ell,\quad
\ell=1,\ldots,L \label{feasible_set_outer_II}
\end{align}
and $\log^+x=\max(\log x,0)$.
\end{Theo}

We obtain this outer bound by evaluating the outer bound given
in~\cite{Wagner_Outer_Bound}. The proof of
Theorem~\ref{theorem_outer} is given in
Section~\ref{sec:proof_of_outer_bound}. Next, we provide the
following inner bound for the rate-distortion region
$\mathcal{R}(\bbd)$.
\begin{Theo}
\label{theorem_inner} An inner bound for the rate-distortion
region of the vector Gaussian CEO problem is given by the region
$\mathcal{R}^{i}(\bbd)$ which is described by the union of rate
tuples $(R_1,\ldots,R_L)$ satisfying
\begin{align}
\sum_{\ell\in\mathcal{A}}R_\ell \geq
\frac{1}{2}\log\frac{\left|\left(\bbk_X^{-1}+\sum_{\ell\in\mathcal{A}^c}\bbsigma_\ell^{-1}-\sum_{\ell\in\mathcal{A}^c}
\bbsigma_\ell^{-1}\bbd_\ell\bbsigma_\ell^{-1}\right)^{-1}\right|}
{\left|\left(\bbk_X^{-1}+\sum_{\ell=1}^L
\bbsigma_\ell^{-1}-\sum_{\ell=1}^L
\bbsigma_\ell^{-1}\bbd_\ell\bbsigma_\ell^{-1}\right)^{-1}\right|}
+\sum_{\ell\in\mathcal{A}}\frac{1}{2}
\log\frac{|\bbsigma_\ell|}{|\bbd_\ell|} \label{rate_bound_inner}
\end{align}
for all $\mathcal{A}\subseteq \{1,\ldots,L\}$, where the union is
over all positive semi-definite matrices
$\{\bbd_\ell\}_{\ell=1}^L$ satisfying
\begin{align}
\left(\bbk_X^{-1}+\sum_{\ell=1}^L\bbsigma_\ell^{-1}-\sum_{\ell=1}^L\bbsigma_\ell^{-1}\bbd_\ell\bbsigma_\ell^{-1}\right)^{-1}&\preceq
\bbd  \label{feasible_set_inner_I}\\
\bzero \preceq \bbd_\ell &\preceq \bbsigma_\ell,\quad
\ell=1,\ldots,L \label{feasible_set_inner_II}
\end{align}
\end{Theo}

We obtain this inner bound by evaluating the Berger-Tung inner
bound~\cite{thesis_tung} by jointly Gaussian auxiliary random
variables. The proof of Theorem~\ref{theorem_inner} is given in
Appendix~\ref{sec:proof_of_inner_bound}.

\section{Alternative Characterizations of the Bounds}

In this section, we provide alternative characterizations for the
outer and inner bounds given in Theorem~\ref{theorem_outer} and
Theorem~\ref{theorem_inner}, respectively. To this end, we note
that since the rate-distortion region $\mathcal{R}(\bbd)$ is
convex, it can be characterized by the tangent hyperplanes to it,
i.e., by solving the following optimization problem
\begin{align}
\min_{(R_1,\ldots,R_L)\in\mathcal{R}(\bbd)}~\sum_{\ell=1}^L~\mu_\ell
R_\ell \label{tangent_lines_original}
\end{align}
for all $\mu_\ell\geq 0,\ell=1,\ldots,L$. Hence, the outer and
inner bounds in Theorem~\ref{theorem_outer} and
\ref{theorem_inner} provide lower and upper bounds for the
optimization problem in (\ref{tangent_lines_original}),
respectively. Since both the outer and inner bounds are also
convex, they can also be described by the tangent hyperplanes to
them. In particular, the outer and inner bounds can be described
by the following optimization problems
\begin{align}
\min_{(R_1,\ldots,R_L)\in\mathcal{R}^o(\bbd)}~\sum_{\ell=1}^L\mu_\ell
R_\ell\qquad \textrm{and}\qquad
\min_{(R_1,\ldots,R_L)\in\mathcal{R}^i(\bbd)}~\sum_{\ell=1}^L\mu_\ell
R_\ell \label{tangent_lines_upper_and_lower}
\end{align}
respectively, where $\mu_\ell\geq 0,~\ell=1,\ldots,L$. We note
that the first optimization problem in
(\ref{tangent_lines_upper_and_lower}) corresponds to the
alternative characterization of the outer bound in
Theorem~\ref{theorem_outer}, and hence, provides a lower bound for
the optimization problem in (\ref{tangent_lines_original}) that
characterizes the rate-distortion region of the vector Gaussian
CEO problem. Similarly, the second optimization problem in
(\ref{tangent_lines_upper_and_lower}) corresponds to the
alternative characterization of the inner bound in
Theorem~\ref{theorem_inner}, and hence, provides an upper bound
for the optimization problem in (\ref{tangent_lines_original}).
Now, we state the explicit form of the optimization problems in
(\ref{tangent_lines_upper_and_lower}) starting with the one for
the outer bound.
\begin{Theo}
\label{theorem_outer_tangent} Assume $\mu_1\geq\ldots\geq
\mu_L\geq 0$. We have
\begin{align}
\lefteqn{\min_{(R_1,\ldots,R_L)\in\mathcal{R}(\bbd)}~\sum_{\ell=1}^L~\mu_\ell
R_\ell \geq
\min_{(R_1,\ldots,R_L)\in\mathcal{R}^o(\bbd)}~\sum_{\ell=1}^L\mu_\ell
R_\ell}\nonumber\\
&=
\min_{\{\bbd_{\ell}\}_{\ell=1}^L}~~\sum_{\ell=1}^{L-1}\frac{\mu_\ell-\mu_{\ell+1}}{2}
\log^+\frac{\left|\left(\bbk_{X}^{-1}+\sum_{j=\ell+1}^L\bbsigma_j^{-1}(\bbsigma_j-\bbd_j)\bbsigma_j^{-1}\right)^{-1}\right|}{|\bbd|}
+\sum_{\ell=1}^L\frac{\mu_\ell}{2}\log\frac{|\bbsigma_\ell|}{|\bbd_\ell|}\nonumber\\
&\qquad\qquad \quad +\frac{\mu_L}{2}\log\frac{|\bbk_X|}{|\bbd|}
\label{rate_bound_outer_tangent}
\end{align}
where $\{\bbd_\ell\}_{\ell=1}^L$ are subject to the following
constraints
\begin{align}
\left(
\bbk_X^{-1}+\sum_{\ell=1}^L\bbsigma_\ell^{-1}-\sum_{\ell=1}^{L}
\bbsigma_\ell^{-1} \bbd_\ell\bbsigma_\ell^{-1}\right)^{-1} &\preceq \bbd \label{feasible_set_outer_I_tangent}\\
\bzero \preceq \bbd_\ell &\preceq \bbsigma_\ell,\quad
\ell=1,\ldots,L \label{feasible_set_outer_II_tangent}
\end{align}
\end{Theo}

Next, we provide the explicit form of the other optimization
problem in (\ref{tangent_lines_upper_and_lower}), i.e., the one
for the inner bound, as follows.
\begin{Theo}
\label{theorem_inner_tangent} Assume $\mu_1\geq\ldots\geq
\mu_L\geq 0$. We have
\begin{align}
\lefteqn{\min_{(R_1,\ldots,R_L)\in\mathcal{R}(\bbd)}~\sum_{\ell=1}^L~\mu_\ell
R_\ell \leq
\min_{(R_1,\ldots,R_L)\in\mathcal{R}^i(\bbd)}~\sum_{\ell=1}^L\mu_\ell
R_\ell}\nonumber\\
&=
\min_{\{\bbd_{\ell}\}_{\ell=1}^L}~~\sum_{\ell=1}^{L-1}\frac{\mu_\ell-\mu_{\ell+1}}{2}
\log\frac{\left|\left(\bbk_{X}^{-1}+\sum_{j=\ell+1}^L\bbsigma_j^{-1}(\bbsigma_j-\bbd_j)\bbsigma_j^{-1}\right)^{-1}\right|}
{\left|\left(\bbk_{X}^{-1}+\sum_{j=1}^L\bbsigma_j^{-1}(\bbsigma_j-\bbd_j)\bbsigma_j^{-1}\right)^{-1}\right|}
+\sum_{\ell=1}^L\frac{\mu_\ell}{2}\log\frac{|\bbsigma_\ell|}{|\bbd_\ell|}\nonumber\\
&\qquad\qquad \quad
+\frac{\mu_L}{2}\log\frac{|\bbk_X|}{\left|\left(\bbk_X^{-1}+\sum_{j=1}^L\bbsigma_j^{-1}(\bbsigma_j-\bbd_j)\bbsigma_j^{-1}\right)^{-1}\right|}
\label{rate_bound_inner_tangent}
\end{align}
where $\{\bbd_\ell\}_{\ell=1}^L$ are subject to the following
constraints
\begin{align}
\left(
\bbk_X^{-1}+\sum_{\ell=1}^L\bbsigma_\ell^{-1}-\sum_{\ell=1}^{L}
\bbsigma_\ell^{-1} \bbd_\ell\bbsigma_\ell^{-1}\right)^{-1} &\preceq \bbd \label{feasible_set_inner_I_tangent}\\
\bzero \preceq \bbd_\ell &\preceq \bbsigma_\ell,\quad
\ell=1,\ldots,L \label{feasible_set_inner_II_tangent}
\end{align}
\end{Theo}

The proofs of Theorem~\ref{theorem_outer_tangent} and
Theorem~\ref{theorem_inner_tangent} are given in
Appendix~\ref{proof_of_tangent_line_bounds}.

Next, we provide some remarks about the outer bound given in
Theorem~\ref{theorem_outer_tangent} and the inner bound given in
Theorem~\ref{theorem_inner_tangent}. First, we note that in both
cases, the bounds are to be optimized over the positive
semi-definite matrices $\{\bbd_\ell\}_{\ell=1}^L$, and the
feasible sets for both cases are identical as seen through
(\ref{feasible_set_outer_I_tangent})-(\ref{feasible_set_outer_II_tangent})
and
(\ref{feasible_set_inner_I_tangent})-(\ref{feasible_set_inner_II_tangent}).
On the other hand, rate bounds differ as seen through
(\ref{rate_bound_outer_tangent}) and
(\ref{rate_bound_inner_tangent}). Despite this difference, there
are cases where the outer and inner bounds match, providing a
complete characterization of the rate-distortion region. Here, we
note a general {\it sufficient} condition under which the outer
and inner bounds coincide. If the minimum in
Theorem~\ref{theorem_outer_tangent} is achieved by positive
semi-definite matrices $\{\bbd^*_\ell\}_{\ell=1}^L$ which attain
the distortion constraint in (\ref{feasible_set_outer_I_tangent})
with equality, then the optimization problems in
Theorem~\ref{theorem_outer_tangent} and
Theorem~\ref{theorem_inner_tangent} yield identical results,
implying the tightness of the outer bound. One particular example
where the outer and inner bounds match is the scalar Gaussian
model considered next.

\subsection{Scalar Gaussian Model}
\label{sec:scalar_Gaussian_model}

In this section, we consider the case where the source and the
observations are scalar:
\begin{align}
Y_{\ell,i}=X_i+N_{\ell,i},\quad \ell=1,\ldots,L
\end{align}
where $X_i$ is an i.i.d. Gaussian source with zero-mean and
variance $\sigma_X^2$. The noise at the $\ell$th sensor
$N_{\ell,i}$ is also an i.i.d. Gaussian random variable sequence
with variance $\sigma_\ell^2$. For the scalar model (scalar
Gaussian CEO problem), our outer bound in
Theorem~\ref{theorem_outer} reduces to the following form.
\begin{Cor}
\label{cor_scalar_outer} The rate-distortion region of the scalar
Gaussian CEO problem $\mathcal{R}(D)$ is contained in the region
$\mathcal{R}^{o}(D)$ which is given by the union of rate tuples
$(R_1,\ldots,R_L)$ satisfying
\begin{align}
\sum_{\ell\in\mathcal{A}}R_\ell\geq \frac{1}{2} \log^+\frac{1}{D}
\left(\frac{1}{\sigma_X^{2}}+\sum_{\ell\in\mathcal{A}^c}\frac{\sigma_\ell^2-D_\ell}{\sigma_\ell^4}\right)^{-1}
+\sum_{\ell\in\mathcal{A}}\frac{1}{2}
\log\frac{\sigma^2_\ell}{D_\ell} \label{rate_bound_outer_scalar}
\end{align}
for all $\mathcal{A}\subseteq\{1,\ldots,L\}$, where the union is
over all $\{D_\ell\}_{\ell=1}^L$ satisfying the following
constraints
\begin{align}
\left(\frac{1}{\sigma_X^{2}}+\sum_{\ell=1}^L\frac{\sigma_\ell^2-D_\ell}{\sigma_\ell^4}\right)^{-1}&\leq D \label{feasible_set_outer_I_scalar_pre}\\
0 \leq D_\ell &\leq \sigma^2_\ell,\quad \ell=1,\ldots,L
\label{feasible_set_outer_II_scalar}
\end{align}
\end{Cor}

Using Theorem~\ref{theorem_outer_tangent}, our outer bound for the
scalar Gaussian model can be expressed in the following
alternative form
\begin{align}
\lefteqn{\min_{(R_1,\ldots,R_L)\in\mathcal{R}(\bbd)}~\sum_{\ell=1}^L~\mu_\ell
R_\ell \geq
\min_{(R_1,\ldots,R_L)\in\mathcal{R}^o(\bbd)}~\sum_{\ell=1}^L\mu_\ell
R_\ell}\nonumber\\
&=\min_{\{D_\ell\}_{\ell=1}^L}~~\sum_{\ell=1}^{L-1}\frac{\mu_\ell-\mu_{\ell+1}}{2}
\log^+\frac{1}{D}\left(\frac{1}{\sigma_X^2}+\sum_{j=\ell+1}^L\frac{\sigma_\ell^2-D_\ell}{\sigma_\ell^4}\right)^{-1}
+\sum_{\ell=1}^L\frac{\mu_\ell}{2}\log\frac{\sigma_\ell^2}{D_\ell}+\frac{\mu_L}{2}\log\frac{\sigma_X^2}{D}
\label{tangent_lines_scalar}
\end{align}
where $\{D_\ell\}_{\ell=1}^L$ are subject to the constraints in
(\ref{feasible_set_outer_I_scalar_pre})-(\ref{feasible_set_outer_II_scalar}),
and we assume $\mu_1\geq \ldots \geq\mu_L \geq 0$.
In~\cite{Oohama_CEO}, it is shown that the optimal
$\{D_\ell^*\}_{\ell=1}^L$ that minimizes
(\ref{tangent_lines_scalar}) satisfies the constraint in
(\ref{feasible_set_outer_I_scalar_pre}) with equality, i.e., for
this optimal $\{D_\ell^*\}_{\ell=1}^L$, we have
\begin{align}
\left(\frac{1}{\sigma_X^{2}}+\sum_{\ell=1}^L\frac{\sigma_\ell^2-D_\ell^*}{\sigma_\ell^4}\right)^{-1}&=
D \label{equality}
\end{align}
As we pointed out in the previous section, when, for the outer
bound, the distortion constraint is satisfied with equality, then
the outer bound in Theorem~\ref{theorem_outer} and the inner bound
in Theorem~\ref{theorem_inner} match; yielding the rate-distortion
region. Hence, in view of (\ref{equality}), we have the entire
rate-distortion region for the scalar Gaussian CEO problem.
\begin{Theo}{\bf (\!\!\cite{Oohama_CEO,Prabhakaran_CEO})}
\label{theo_scalar} The rate-distortion region of the scalar
Gaussian CEO problem $\mathcal{R}(D)$ is given by the union of
rate tuples $(R_1,\ldots,R_L)$ satisfying
\begin{align}
\sum_{\ell\in\mathcal{A}}R_\ell\geq \frac{1}{2} \log\frac{1}{D}
\left(\frac{1}{\sigma_X^{2}}+\sum_{\ell\in\mathcal{A}^c}\frac{\sigma_\ell^2-D_\ell}{\sigma_\ell^4}\right)^{-1}
+\sum_{\ell\in\mathcal{A}}\frac{1}{2}
\log\frac{\sigma^2_\ell}{D_\ell} \label{rate_bounds_scalar_final}
\end{align}
for all $\mathcal{A}\subseteq\{1,\ldots,L\}$, where the union is
over all $\{D_\ell\}_{\ell=1}^L$ satisfying the following
constraints
\begin{align}
\left(\frac{1}{\sigma_X^{2}}+\sum_{\ell=1}^L\frac{\sigma_\ell^2-D_\ell}{\sigma_\ell^4}\right)^{-1}&= D \label{feasible_set_outer_I_scalar}\\
0 \leq D_\ell &\leq \sigma^2_\ell,\quad \ell=1,\ldots,L
\end{align}
\end{Theo}

We note that since the distortion constraint in
(\ref{feasible_set_outer_I_scalar}) is satisfied with equality, we
do not need the positivity operator in
(\ref{rate_bounds_scalar_final}).

\section{Chen-Wang Outer Bound}

\label{sec:Chen-Wang}

In~\cite[Theorem~2]{Chen_Wang_Vector_CEO}, the authors provide an
outer bound for the rate-distortion region of the vector Gaussian
CEO problem when $L=2$. In this section, we compare our outer
bound given in Theorem~\ref{theorem_outer}. First, we note that
the outer bound in~\cite[Theorem~2]{Chen_Wang_Vector_CEO} always
contains our outer bound for all system parameters. Next, we
provide an example and show that the outer bound
in~\cite[Theorem~2]{Chen_Wang_Vector_CEO} {\it strictly} contains
our outer bound. In other words, we show that there are rate pairs
$(R_1,R_2)$ that are contained in the outer bound given
in~\cite[Theorem~2]{Chen_Wang_Vector_CEO} and are strictly outside
of our outer bound given in Theorem~\ref{theorem_outer}. To this
end, we specialize our outer bound in
Theorem~\ref{theorem_outer_tangent} to the case $L=2$ as follows.
\begin{Cor}
\label{corollary_our_outer_bound} When $\mu_1\geq \mu_2 \geq 0$,
we have
\begin{align}
\mu_1 R_1+\mu_2 R_2\geq {\rm
T^+}&=\min_{(\bbd_1,\bbd_2)\in\mathcal{D}^+(\bbd_1,\bbd_2)}~\frac{\mu_1}{2}
\log\frac{|\bbsigma_1|}{|\bbd_1|}+\frac{\mu_2}{2}\log\frac{|\bbsigma_2|}{|\bbd_2|}
+\frac{\mu_2}{2}\log\frac{|\bbk_X|}{|\bbd|}\nonumber \\
&\qquad \qquad \qquad \qquad  + \frac{\mu_1-\mu_2}{2}
\log^{+}\frac{\left|\left(\bbk_X^{-1}+\bbsigma_2^{-1}-\bbsigma_2^{-1}\bbd_2\bbsigma_2^{-1}\right)^{-1}\right|}{|\bbd|}
\label{our_outer_bound_1}
\end{align}
where the feasible set $\mathcal{D}^+(\bbd_1,\bbd_2)$ is given by
the union of $(\bbd_1,\bbd_2)$ satisfying
\begin{align}
\left(\bbk_X^{-1}+\sum_{\ell=1}^2\bbsigma_\ell^{-1}-\sum_{\ell=1}^2\bbsigma_\ell^{-1}\bbd_\ell\bbsigma_\ell^{-1}\right)^{-1}&\preceq
\bbd \label{our_outer_bound_2} \\
\bzero\preceq \bbd_\ell&\preceq \bbsigma_\ell,~\quad \ell=1,2
\label{our_outer_bound_3}
\end{align}
When $0\leq \mu_1\leq \mu_2$, a lower bound for $\mu_1 R_1+\mu_2
R_2$ can be obtained from
(\ref{our_outer_bound_1})-(\ref{our_outer_bound_3}) by swapping
the indices $1$ and $2$.
\end{Cor}

Now, we present the outer bound
in~\cite[Theorem~2]{Chen_Wang_Vector_CEO}.
\begin{Theo}{\bf(\!\!\cite[Theorem~2]{Chen_Wang_Vector_CEO})}
\label{theorem_outer_chen} When $\mu_1\geq \mu_2 \geq 0$, we have
\begin{align}
\mu_1 R_1+\mu_2 R_2\geq {\rm
T^-}&=\min_{(\bbd_1,\bbd_2)\in\mathcal{D}^-(\bbd_1,\bbd_2)}~\frac{\mu_1}{2}
\log\frac{|\bbsigma_1|}{|\bbd_1|}+\frac{\mu_2}{2}\log\frac{|\bbsigma_2|}{|\bbd_2|}
+\frac{\mu_2}{2}\log\frac{|\bbk_X|}{|\bbd|}\nonumber \\
&\qquad \qquad \qquad  \qquad + \frac{\mu_1-\mu_2}{2}
\log\frac{\left|\left(\bbk_X^{-1}+\bbsigma_2^{-1}-\bbsigma_2^{-1}\bbd_2\bbsigma_2^{-1}\right)^{-1}\right|}{|\bbd|}
\label{chen_outer_bound_1}
\end{align}
where the feasible set $\mathcal{D}^-(\bbd_1,\bbd_2)$ is given by
the union of $(\bbd_1,\bbd_2)$ satisfying
\begin{align}
\left(\bbk_X^{-1}+\sum_{\ell=1}^2\bbsigma_\ell^{-1}-\sum_{\ell=1}^2\bbsigma_\ell^{-1}\bbd_\ell\bbsigma_\ell^{-1}\right)^{-1}&\preceq
\bbd \label{chen_outer_bound_2} \\
\bzero\preceq \bbd_\ell&\preceq \bbsigma_\ell,~\quad \ell=1,2
\label{chen_outer_bound_3}
\end{align}
When $0\leq \mu_1\leq \mu_2$, a lower bound for $\mu_1 R_1+\mu_2
R_2$ can be obtained from
(\ref{chen_outer_bound_1})-(\ref{chen_outer_bound_3}) by swapping
the indices $1$ and $2$.
\end{Theo}

We note that the only difference between the outer bounds in
Corollary~\ref{corollary_our_outer_bound} and
Theorem~\ref{theorem_outer_chen} is the positivity operator
involved in (\ref{our_outer_bound_1}) (compare
(\ref{our_outer_bound_1}) with (\ref{chen_outer_bound_1})).
Besides that, the two outer bounds are identical. In the sequel,
we first provide an outline for both approaches that explains how
the difference between these two outer bounds arises. We note that
because of the positivity operator in our outer bound, we always
have ${\rm T^+}\geq {\rm T^-}$ in general, and our outer bound is
at least as tight as the outer bound
in~\cite[Theorem~2]{Chen_Wang_Vector_CEO} or tighter, for all
instances of the vector Gaussian CEO problem. Next, we provide an
example where ${\rm T^+}>{\rm T^-}$, which implies that our outer
bound is strictly contained in the outer bound given
in~\cite[Theorem~2]{Chen_Wang_Vector_CEO}.

In~\cite{Chen_Wang_Vector_CEO}, the lower bound ${\rm T^-}$ is
obtained by minimizing the following cost function
\begin{align}
{\rm C^-}&=\frac{\mu_1}{n}I(B_1^n;\bby_1^n|\bbx^n)+\frac{\mu_2}{n}
I(\bbx^n;B_1^n,B_2^n)+\frac{\mu_1-\mu_2}{n} I(\bbx^n;B_1^n|B_2^n)
+\frac{\mu_2}{n} I(\bby_2^n;B_2^n|\bbx^n)
\label{cost_function_chen}
\end{align}
where the authors consider the first and the second terms
separately, which leads to the following terms
\begin{align}
\frac{\mu_1}{2} \log\frac{|\bbsigma_1|}{|\bbd_1|}\quad {\rm
and}\quad \frac{\mu_2}{2}\log\frac{|\bbk_X|}{|\bbd|}
\end{align}
in (\ref{chen_outer_bound_1}), respectively. The third and fourth
terms in (\ref{cost_function_chen}) are considered jointly. In
particular, in~~\cite[Theorem~2]{Chen_Wang_Vector_CEO}, the
authors rewrite the third and fourth terms as
\begin{align}
{\rm C_{3,4}^-}=\left[\frac{\mu_1-\mu_2}{2n} h(\bbx^n|B_2^n)
-\frac{\mu_2}{2n}
h(\bby_2^n|B_2^n,\bbx^n)\right]-\frac{\mu_1-\mu_2}{2n}
h(\bbx^n|B_1^n,B_2^n)+\frac{\mu_2}{2n} h(\bby_2^n|\bbx^n)
\label{sub_cost_function_chen}
\end{align}
and minimize ${\rm C_{3,4}^-}$. In particular, the difference term
in the bracket is minimized jointly, which is the reason why there
is no positivity operator in the outer bound given by
Theorem~\ref{theorem_outer_chen}. On the other hand, we consider
the following cost function
\begin{align}
{\rm C^+}=\mu_1 I(U_1;\bby_1|\bbx,W)+\mu_2
I(\bbx;U_1,U_2)+(\mu_1-\mu_2)I(\bbx;U_1|U_2)+\mu_2
I(\bby_2;U_2|\bbx,W)
\end{align}
which can be obtained by using the outer bound provided
in~\cite{Wagner_Outer_Bound}. (More details about the cost
function ${\rm C^{+}}$ can be found in
Section~\ref{sec:proof_of_outer_bound}, where we prove
Theorem~\ref{theorem_outer}.) We note that the cost function ${\rm
C^+}$ can be viewed as a single-letter form of the cost function
${\rm C^-}$. As opposed to~\cite{Chen_Wang_Vector_CEO} where the
mutual information terms involved in the cost function ${\rm
C^{-}}$ are decomposed into differential entropies and some cross
terms are minimized jointly (see ${\rm C_{3,4}^-}$), we consider
each mutual information term in the cost function ${\rm C^+}$
separately, and find a lower bound for each term. Hence, we find a
lower bound for the third term in ${\rm C^+}$ which, being a
mutual information, is non-negative. This is the reason why we
have a positivity operator in our outer bound given in
Corollary~\ref{corollary_our_outer_bound} (and also in Theorem
~\ref{theorem_outer} and Theorem~\ref{theorem_outer_tangent}).

Next, we provide an example where we have ${\rm T^+}>{\rm T^-}$,
which implies that our outer bound in
Corollary~\ref{corollary_our_outer_bound} (and, hence in
Theorem~\ref{theorem_outer}) is strictly contained in the outer
bound~\cite[Theorem~2]{Chen_Wang_Vector_CEO} in
Theorem~\ref{theorem_outer_chen}. In other words, there are rate
pairs $(R_1,R_2)$ that lie inside the outer bound given by
Theorem~\ref{theorem_outer_chen} and lie strictly outside of our
outer bound. To show this, we consider the case where the
following assumptions hold:
\begin{align}
\frac{\mu_2}{\mu_1}\bbsigma_1^{-1}&\prec
\bbk_X^{-1}+\bbsigma_2^{-1}-\bbd^{-1} \label{first_assumption} \\
\frac{\mu_2}{\mu_1-\mu_2}\bbk_X^{-1} &\prec \bbsigma_2^{-1}
\label{second_assumption}\\
\frac{\mu_1}{\mu_1-\mu_2} \bbd^{-1}&\prec
\bbk_X^{-1}+\bbsigma_2^{-1} \label{third_assumption}
\end{align}
Under the assumptions given by
(\ref{first_assumption})-(\ref{third_assumption})\footnote{An
example where these conditions hold is
$\bbk_X^{-1}=\bbsigma_1^{-1}=\bbsigma_2^{-1}$ and $\mu_1/\mu_2=4$.
For this case, one can find $\bbd$ matrices satisfying these
constraints in addition to the original constraints on $\bbd$
stated in (\ref{limits_of_distortion}).}, we can obtain our outer
bound given in Corollary~\ref{corollary_our_outer_bound}
explicitly in terms of $\bbk_X,\bbd,\bbsigma_\ell$ and
$\mu_\ell,~\ell=1,2,$ as stated in the following corollary.
\begin{Cor}
\label{Corollary_after_assumptions} When the assumptions given by
(\ref{first_assumption})-(\ref{third_assumption}) hold, we have
\begin{align}
{\rm T^+}=\frac{\mu_2}{2}\log\frac{|\bbk_X|}{|\bbd|}+
\frac{\mu_2}{2}\log\frac{|\bbsigma_2^{-1}|}{|\bbk_X^{-1}+\bbsigma_2^{-1}-\bbd^{-1}|}
\end{align}
\end{Cor}

Next, we obtain an upper bound for the lower bound given in
Theorem~\ref{theorem_outer_chen}. In other words, we obtain an
upper bound for ${\rm T^{-}}$ as stated in the following
corollary.
\begin{Cor}
\label{Corollary_after_assumptions_chen} When the assumptions
given by (\ref{first_assumption})-(\ref{third_assumption}) hold,
we have
\begin{align}
{\rm T^-}\leq {\rm T^+}+ \frac{\mu_2}{2}
\log\frac{|\bbk_X^{-1}+\bbsigma_2^{-1}-\bbd^{-1}|}{\left|\frac{\mu_2}{\mu_1}\left(\bbk_X^{-1}+\bbsigma_2^{-1}\right)\right|}
+\frac{\mu_1-\mu_2}{2}\log\frac{\left|\frac{\mu_1}{\mu_1-\mu_2}\left(\bbk_X^{-1}+\bbsigma_2^{-1}\right)^{-1}\right|}{|\bbd|}
\end{align}
\end{Cor}

The proofs of Corollaries~\ref{Corollary_after_assumptions}
and~\ref{Corollary_after_assumptions_chen} are given in
Appendix~\ref{sec:proof_of_corollary_after_assumptions} and
Appendix~\ref{sec:proof_of_corollary_after_assumptions_chen},
respectively.

Now, we are ready to compare ${\rm T^+}$ and ${\rm T^-}$ as
follows
\begin{align}
{\rm T^-}-{\rm T^+}&\leq
\frac{\mu_2}{2}\log\frac{|\bbk_X^{-1}+\bbsigma_2^{-1}-\bbd^{-1}|}
{\left|\frac{\mu_2}{\mu_1}\left(\bbk_X^{-1}+\bbsigma_2^{-1}\right)\right|}
+\frac{\mu_1-\mu_2}{2}\log\frac{\left|\frac{\mu_1}{\mu_1-\mu_2}\left(\bbk_X^{-1}+\bbsigma_2^{-1}\right)^{-1}\right|}{|\bbd|}
\\
&=\frac{\mu_2}{2}\log\left|\frac{\mu_1}{\mu_2}\left(\bbi-\left(\bbk_X^{-1}+\bbsigma_2^{-1}\right)^{-1/2}\bbd^{-1}\left(\bbk_X^{-1}+\bbsigma_2^{-1}\right)^{-1/2}\right)\right|
\nonumber\\
&\quad +
\frac{\mu_1-\mu_2}{2}\log\left|\frac{\mu_1}{\mu_1-\mu_2}\left(\bbk_X^{-1}+\bbsigma_2^{-1}\right)^{-1/2}\bbd^{-1}\left(\bbk_X^{-1}+\bbsigma_2^{-1}\right)^{-1/2}\right|
\label{not_identical}\\
&<\frac{\mu_1}{2}\log|\bbi|\label{strict_concavity_implies}\\
&=0
\end{align}
where (\ref{strict_concavity_implies}) follows from the facts that
the function $\log|\cdot|$ is strictly concave over strictly
positive definite matrices~\cite[Theorem
7.6.7]{horn_johnson_book1}, and the two matrices inside the
$\log|\cdot|$ functions in (\ref{not_identical}) are not
identical, which is due to the assumption in
(\ref{third_assumption}).

\section{Parallel Gaussian Model and a Counter-Example}

In this section, first, we consider the parallel Gaussian model,
and obtain its rate-distortion region. Next, we consider a
specific parallel Gaussian model and show that our outer bound in
Theorem~\ref{theorem_outer} is not tight. In other words, we show
that, in general, there are rate tuples $(R_1,\ldots,R_L)$ that
lie inside our outer bound and are not contained in the
rate-distortion region, i.e., in general, our outer bound strictly
contains the rate-distortion region.

 In the parallel Gaussian model, the Gaussian
source $\bbx_i$ has a diagonal covariance matrix. In particular,
we have $\bbx_i=[~X_{1,i}~\ldots~X_{M,i}~]$ where
$\{X_{m,i}\}_{m=1}^M$ are independent Gaussian random variables
with zero-mean and variance $\{\sigma_m^2\}_{m=1}^M$,
respectively. Moreover, the noise at the $\ell$th sensor
$\bbn_{\ell,i}$ also has a diagonal covariance matrix. In
particular, we have $\bbn_{\ell,i}=[~N_{\ell 1,i}~\ldots~N_{\ell
M,i}~]$, where $\{N_{\ell m,i}\}_{m=1}^M$ are independent Gaussian
random variables with zero-mean with variance $\{\sigma_{\ell
m}^2\}_{m=1}^M$, respectively. In the parallel Gaussian model,
there is a separate-distortion constraint on each component of the
source as follows
\begin{align}
\lim_{n\rightarrow \infty}~\frac{1}{n}\sum_{i=1}^n
\mmse(X_{m,i}|B_1^n,\ldots,B_L^n)\leq D_m,~\quad m=1,\ldots,M
\end{align}
where we have the following constraints on $\{D_m\}_{m=1}^M$
\begin{align}
\left(\frac{1}{\sigma_m^2}+\sum_{\ell=1}^L\frac{1}{\sigma_{\ell
m}^2}\right)^{-1}\leq D_m \leq \sigma_m^2,\quad m=1,\ldots,M
\label{limits_of_distortion_scalar}
\end{align}
We note that the constraints on $D_m$ in
(\ref{limits_of_distortion_scalar}) are the scalar versions of the
constraints in (\ref{limits_of_distortion}) that we impose for the
vector Gaussian model. For the parallel Gaussian model, we
establish the rate-distortion region $\mathcal{R}^{
p}(\{D_m\}_{m=1}^M)$ as stated in the following theorem.
\begin{Theo}
\label{theorem_r_d_parallel} The rate-distortion region
$\mathcal{R}^{p}(\{D_m\}_{m=1}^M)$ of the parallel Gaussian CEO
problem is given by the union of rate tuples $(R_1,\ldots,R_L)$
satisfying
\begin{align}
\sum_{\ell\in\mathcal{A}}R_\ell&\geq
\sum_{m=1}^M~\frac{1}{2}\log\frac{1}{D_m}\left(\frac{1}{\sigma_m^2}+\sum_{\ell\in\mathcal{A}^c}\frac{\sigma_{\ell
m}^2-D_{\ell m}}{\sigma_{\ell m}^4}\right)^{-1} +\sum_{m=1}^M
\sum_{\ell\in\mathcal{A}} \frac{1}{2}\log\frac{\sigma_{\ell
m}^2}{D_{\ell m}}\label{rate_bound_parallel}
\end{align}
for all $\mathcal{A}\subseteq \{1,\ldots,L\}$, where the union is
over all $\{D_{\ell m}\}_{\forall \ell, \forall m}$ satisfying the
following constraints
\begin{align}
\left(\frac{1}{\sigma_m^2}+\sum_{\ell=1}^L \frac{\sigma_{\ell
m}^2-D_{\ell m}}{\sigma_{\ell m}^4}\right)^{-1}&=D_m,\quad
m=1,\ldots,M \label{distortion_constraint_parallel_x} \\
0\leq D_{\ell m}&\leq \sigma_{\ell m}^2,\quad
\ell=1,\ldots,L,~~m=1,\ldots,M
\end{align}
\end{Theo}

We note that since the distortion constraints in
(\ref{distortion_constraint_parallel_x}) are met with equality,
the first $\log (\cdot)$ in (\ref{rate_bound_parallel}) is always
positive, and hence, we do not need a positivity operator. We
obtain the rate-distortion region of the parallel Gaussian CEO
problem in two steps. In the first step, we specialize the outer
bound in~\cite{Wagner_Outer_Bound} to the parallel model. In the
second step, we evaluate the outer bound we obtain in the first
step, and show that it matches the inner bound given in
Theorem~\ref{theorem_inner}. The details of the proof are given in
Appendix~\ref{proof_of_theorem_r_d_parallel}.

Next, we consider the case $L=M=2$, and provide an example where
our outer bound strictly contains the rate-distortion region,
i.e., our outer bound includes rate pairs which are outside of the
rate-distortion region. In the example we provide, we assume that
the following conditions hold\footnote{We note that if one selects
$\sigma_m^2=\sigma_{\ell m}^2=\sigma^2,
D_1=2/5\sigma^2,D_2=4/5\sigma^2$ and $\mu_1/\mu_2=4$, the four
assumptions in
(\ref{first_assumption_p})-(\ref{fourth_assumption_p}) hold in
addition to the original constraints on $(D_1,D_2)$ given in
(\ref{limits_of_distortion_scalar}).}:
\begin{align}
\frac{\mu_2}{\mu_1}\frac{1}{\sigma_{12}^2}&<\frac{1}{\sigma_2^2}+\frac{1}{\sigma_{22}^2}-\frac{1}{D_2}
\label{first_assumption_p}\\
\frac{\mu_2}{\mu_1-\mu_2}\frac{1}{\sigma_2^2}&<\frac{1}{\sigma_{22}^2}
\label{second_assumption_p}\\
\frac{\mu_1}{\mu_1-\mu_2}\frac{1}{D_2}&<\frac{1}{\sigma_2^2}+\frac{1}{\sigma_{22}^2}
\label{third_assumption_p}\\
\frac{1}{D_1}\left(\frac{1}{\sigma_1^2}+\frac{1}{\sigma_{21}^2}\right)^{-1}&>\frac{\mu_1-\mu_2}{\mu_1}D_2
\left(\frac{1}{\sigma_2^2}+\frac{1}{\sigma_{22}^2}\right)
\label{fourth_assumption_p}
\end{align}
where the first three constraints are analogous the constraints in
(\ref{first_assumption})-(\ref{third_assumption}), which were used
to provide an example that the Chen-Wang outer
bound~\cite{Chen_Wang_Vector_CEO} strictly contains our outer
bound. Under the constraints in
(\ref{first_assumption_p})-(\ref{fourth_assumption_p}), the
rate-distortion region $\mathcal{R}^p(D_1,D_2)$ can be
characterized as follows.
\begin{Cor}
\label{Corollary_after_assumptions_p} Assume that
(\ref{first_assumption_p})-(\ref{fourth_assumption_p}) hold. Then,
we have
\begin{align}
{\rm T}^p&=\min_{(R_1,R_2)\in\mathcal{R}^p(D_1,D_2)}~\mu_1
R_1+\mu_2 R_2\\
&=\min_{(D_{11},D_{21})\in\mathcal{D}_1}f_1
(D_{11},D_{21})+\frac{\mu_2}{2}\log\frac{\sigma_2^2}{D_2}
+\frac{\mu_2}{2}
\log\frac{1}{\sigma_{22}^2}\left(\frac{1}{\sigma_2^2}+\frac{1}{\sigma_{22}^2}-\frac{1}{D_2}\right)^{-1}
\end{align}
where $f_{1}(D_{11},D_{21})$ is given by
\begin{align}
f_{1}(D_{11},D_{21})&=\sum_{\ell=1}^2\frac{\mu_\ell}{2}
\log\frac{\sigma_{\ell 1}^2}{D_{\ell 1}}+\frac{\mu_2}{2}
\log\frac{\sigma_1^2}{D_1}
+\frac{\mu_1-\mu_2}{2}\log\frac{1}{D_1}\left(\frac{1}{\sigma_1^2}+\frac{\sigma_{21}^2-D_{21}}{\sigma_{21}^4}\right)^{-1}
\label{function_1}
\end{align}
and the set $\mathcal{D}_1$ consists of $(D_{11},D_{21})$ pairs
satisfying
\begin{align}
\frac{1}{\sigma_{1}^2}+\sum_{\ell=1}^2 \frac{\sigma_{\ell
1}^2-D_{\ell 1}}{\sigma_{\ell 1}^4}&=\frac{1}{D_1} \label{D_1_constraint_1}\\
0\leq D_{\ell 1} &\leq \sigma_{\ell 1}^2,\quad \ell=1,2
\label{D_1_constraint_2}
\end{align}
\end{Cor}

The proof of Corollary~\ref{Corollary_after_assumptions_p} is
given in Appendix~\ref{proof_of_Corollary_after_assumptions_p}.
Next, we find an upper bound for our outer bound in
Theorem~\ref{theorem_outer} as follows.
\begin{Cor}
\label{corollary_after_assumptions_ours_p} Assume that
(\ref{first_assumption_p})-(\ref{fourth_assumption_p}) hold. Then,
we have
\begin{align}
{\rm T^+}&=\min_{(R_1,R_2)\in\mathcal{R}^o(D_1,D_2)}~\mu_1
R_1+\mu_2 R_2\\
&\leq \min_{(D_{11},D_{21})\in\mathcal{D}_1}~f_{1}(D_{11},D_{21})
+\frac{\mu_2}{2} \log\frac{\mu_1}{\mu_2}\frac{1}{\sigma_{22}^2}
\left(\frac{1}{\sigma_2^2}+\frac{1}{\sigma_{22}^2}\right)^{-1}
+\frac{\mu_2}{2}\log\frac{\sigma_2^2}{D_2}
\nonumber\\
&\qquad +\frac{\mu_1-\mu_2}{2}
\log\frac{\mu_1}{\mu_1-\mu_2}\frac{1}{D_2}\left(\frac{1}{\sigma_2^2}+\frac{1}{\sigma_{22}^2}\right)^{-1}
\end{align}
where the function $f_1(D_{11},D_{21})$ is given by
(\ref{function_1}) and the set $\mathcal{D}_1$ is given by the
union of $(D_{11},D_{21})$ satisfying the constraints in
(\ref{D_1_constraint_1})-(\ref{D_1_constraint_2}).
\end{Cor}

The proof of Corollary~\ref{corollary_after_assumptions_ours_p} is
given in
Appendix~\ref{proof_of_corollary_after_assumptions_ours_p}.

Now, we are ready to compare our outer bound with the
rate-distortion region for the parallel Gaussian model. Using
Corollary~\ref{Corollary_after_assumptions_p} and
Corollary~\ref{corollary_after_assumptions_ours_p}, we have
\begin{align}
{\rm T^+}-{\rm T}^p&\leq \frac{\mu_2}{2}
\log\frac{\mu_1}{\mu_2}\frac{1}{\sigma_{22}^2}\left(\frac{1}{\sigma_2^2}+\frac{1}{\sigma_{22}^2}\right)^{-1}
+\frac{\mu_1-\mu_2}{2}\log\frac{\mu_1}{\mu_1-\mu_2}\frac{1}{D_2}\left(\frac{1}{\sigma_2^2}+\frac{1}{\sigma_{22}^2}\right)^{-1}\nonumber\\
&\quad
-\frac{\mu_2}{2}\log\frac{1}{\sigma_{22}^2}\left(\frac{1}{\sigma_2^2}+\frac{1}{\sigma_{22}^{2}}-\frac{1}{D_2}\right)^{-1}
\\
&= \frac{\mu_2}{2}
\log\frac{\mu_1}{\mu_2}\left(1-\frac{1}{D_2}\left(\frac{1}{\sigma_2^2}+\frac{1}{\sigma_{22}^2}\right)^{-1}\right)
+\frac{\mu_1-\mu_2}{2}\log\frac{\mu_1}{\mu_1-\mu_2}\frac{1}{D_2}\left(\frac{1}{\sigma_2^2}+\frac{1}{\sigma_{22}^2}\right)^{-1}\\
&<\frac{\mu_1}{2} \log 1 \label{strict_concavity_implies_again}\\
&=0 \label{strict_concavity_implies_again_1}
\end{align}
where (\ref{strict_concavity_implies_again}) follows from the
facts that $\log (\cdot)$ is strictly concave, and we have
\begin{align}
\frac{\mu_1}{\mu_2}\left(1-\frac{1}{D_2}\left(\frac{1}{\sigma_2^2}+\frac{1}{\sigma_{22}^2}\right)^{-1}\right)
\neq
\frac{\mu_1}{\mu_1-\mu_2}\frac{1}{D_2}\left(\frac{1}{\sigma_2^2}+\frac{1}{\sigma_{22}^2}\right)^{-1}
\end{align}
which is due to the assumption in (\ref{third_assumption_p}).
Equation (\ref{strict_concavity_implies_again_1}) implies that
there are some rate pairs $(R_1,R_2)$ in our outer bound which are
outside of the rate-distortion region of the parallel Gaussian
model. Hence, our outer bound strictly contains the
rate-distortion region of the vector Gaussian CEO problem. In
other words, our outer bound is not tight in general.

\section{Proof of Theorem~\ref{theorem_outer}}

\label{sec:proof_of_outer_bound}

The following theorem provides an outer bound for the
rate-distortion region of the CEO problem.

\begin{Theo}{\bf(\!\!\cite[Theorem 1]{Wagner_Outer_Bound})}
The rate region of the CEO problem $\mathcal{R}(\bbd)$ is
contained in the union of rate tuples $(R_1,\ldots,R_L)$
satisfying
\begin{align}
\sum_{\ell\in\mathcal{A}}R_\ell &\geq
I(\bbx;\{U_\ell\}_{\ell\in\mathcal{A}}|\{U_\ell\}_{\ell\in\mathcal{A}^c})+\sum_{\ell\in\mathcal{A}}I(\bby_\ell;U_\ell|\bbx,W),\quad
\forall \mathcal{A}\subseteq \{1,\ldots,L\}
\label{outer_bound_general}
\end{align}
where the union is over all joint distributions
$p(\bx,\{\by_\ell,u_\ell\}_{\ell=1}^L,w)$ that can be factorized
as
\begin{align}
p(\bx,\{\by_\ell,u_\ell\}_{\ell=1}^L,w)=p(\bx)p(w)\prod_{\ell=1}^L
p(\by_\ell|\bx)p(u_\ell|\by_\ell,w) \label{joint_distribution}
\end{align}
and satisfies
\begin{align}
\mmse (\bbx|U_1,\ldots,U_L)&\preceq \bbd
\label{distortion_constraint}
\end{align}
\end{Theo}
In~\cite{Wagner_Outer_Bound}, the outer bound is stated in a
slightly different form, where there is a time-sharing random
variable $T$ involved in the description of the outer bound.
However, as pointed out by~\cite{Wagner_Outer_Bound}, this
time-sharing random variable $T$ can be combined with other
auxiliary random variables $(W,U_1,\ldots,U_L)$ to obtain the form
of the outer bound we stated here.

We now evaluate this outer bound for the vector Gaussian CEO
problem. To this end, we first provide some background information
which will be used in the proof.

\subsection{Background}

\begin{Lem}{\bf (\!\!\cite{MIMO_BC_Secrecy})}
\label{lemma_conditional_crb} Let $(U,\bbx)$ be an arbitrarily
correlated random vector with well-defined densities. We assume
that $\mmse(\bbx|U)\succ \bzero$. Then, we have
\begin{align}
\bbj(\bbx|U)\succeq \mmse^{-1}(\bbx|U)
\end{align}
which is satisfied with equality if $(U,\bbx)$ is jointly
Gaussian.
\end{Lem}

Next, we note the following lemma which will be used subsequently.
\begin{Lem}[\!\!\cite{Dembo,Dembo_Cover}]
\label{lemma_dembo} Let $(U,\bbx)$ be an arbitrary random vector,
where the conditional Fisher information of $\bbx$, conditioned on
$U$, exists. Then, we have
\begin{align}
\frac{1}{2} \log |(2\pi e) \bbj^{-1}(\bbx|U)| \leq h(\bbx|U)
\end{align}
\end{Lem}

We also need the following lemma in the upcoming proof.
\begin{Lem}{\bf (\!\!\cite{Palomar_Gradient})}
\label{lemma_connection} Let $(\bbv_1,\bbv_2)$ be an arbitrary
random vector with finite second moments, and $\bbn$ be a
zero-mean Gaussian random vector with covariance $\bbsigma_N$.
Assume $(\bbv_1,\bbv_2)$ and $\bbn$ are independent. We have
\begin{align}
\mmse(\bbv_2|\bbv_1,\bbv_2+\bbn)=\bbsigma_N-\bbsigma_N
\bbj(\bbv_2+\bbn|\bbv_1) \bbsigma_N
\end{align}
\end{Lem}

\subsection{Proof}

Here, we consider the rate bounds in (\ref{outer_bound_general})
and obtain a lower bound for them for a given
$(W,U_1,\ldots,U_L)$. First, we consider the following mutual
information terms
\begin{align}
I(\bby_\ell;U_\ell|\bbx,W)&=h(\bby_\ell|\bbx,W)-h(\bby_\ell|\bbx,W,U_\ell)\\
&=h(\bby_\ell|\bbx)-h(\bby_\ell|\bbx,W,U_\ell)\\
&=\frac{1}{2}\log|(2\pi e)\bbsigma_\ell
|-h(\bby_\ell|\bbx,W,U_\ell) \label{dummy_1}
\end{align}
Using Lemma~\ref{lemma_dembo} and the fact that jointly Gaussian
$(\bbx,W,U_\ell,\bby_\ell)$ maximizes
$h(\bby_\ell|\bbx,W,U_\ell)$, we have the following bounds for the
second term in (\ref{dummy_1})
\begin{align}
\frac{1}{2}\log |(2\pi e) \bbj^{-1}(\bby_\ell|\bbx,W,U_\ell)|\leq
h(\bby_\ell|\bbx,W,U_\ell) \leq \frac{1}{2}\log |(2\pi
e)\mmse(\bby_\ell|\bbx,W,U_\ell)| \label{dummy_2}
\end{align}
Next, we define the function $\bbd_{\ell}(\alpha_\ell)$ as follows
\begin{align}
\bbd_\ell(\alpha_\ell)=\alpha_\ell
\bbj^{-1}(\bby_\ell|\bbx,W,U_\ell)+\bar{\alpha}_\ell
\mmse(\bby_\ell|\bbx,W,U_\ell) \label{f_1}
\end{align}
where $\alpha_\ell=1-\bar{\alpha}_\ell\in[0,1]$. Using the
function in (\ref{f_1}), the bounds in (\ref{dummy_2}) can be
expressed as follows
\begin{align}
\frac{1}{2}\log |(2\pi e)\bbd_\ell(1)|\leq
h(\bby_\ell|\bbx,W,U_\ell) \leq \frac{1}{2} \log |(2\pi
e)\bbd_\ell(0)|
\end{align}
Since $\log |(2\pi e) \bbd_\ell(\alpha_\ell)|$ is continuous in
$\alpha_\ell$, due to the intermediate value theorem, there exists
an $\alpha^*_\ell=1-\bar{\alpha}_\ell^*\in[0,1]$ such that
\begin{align}
h(\bby_\ell|\bbx,W,U_\ell)&=\frac{1}{2}\log |(2\pi e)\bbd_\ell(\alpha_\ell^*)| \\
&=\frac{1}{2} \log \left|(2\pi e) \left(\alpha^*_\ell
\bbj^{-1}(\bby_\ell|\bbx,W,U_\ell)+\bar{\alpha}^*_\ell
\mmse(\bby_\ell|\bbx,W,U_\ell) \right)\right| \label{fix_1}
\end{align}
Hence, using (\ref{fix_1}) in (\ref{dummy_1}), we have
\begin{align}
I(\bby_\ell;U_\ell|\bbx,W)=\frac{1}{2}\log\frac{|\bbsigma_\ell|}{|\bbd_\ell(\alpha_\ell^*)|},\quad
\ell=1,\ldots,L \label{fixed_ones}
\end{align}
We note the following bounds on $\bbd_\ell(\alpha_\ell^*)$
\begin{align}
\bbj^{-1}(\bby_\ell|\bbx,W,U_\ell)\preceq
\bbd_\ell(\alpha_\ell^*)&\preceq \mmse(\bby_\ell|\bbx,W,U_\ell)
\label{order_1} \\
&\preceq \mmse(\bby_\ell|\bbx)\label{conditioning_reduces_MMSE_X}\\
&=\bbsigma_\ell \label{less_than_noise}
\end{align}
where (\ref{order_1}) is due to Lemma~\ref{lemma_conditional_crb}
and (\ref{conditioning_reduces_MMSE_X}) comes from the fact that
conditioning reduces the MMSE matrix in the positive semi-definite
ordering sense.

Next, we consider the following mutual information term
\begin{align}
I(\bbx;\{U_\ell\}_{\ell\in\mathcal{A}}|\{U_\ell\}_{\ell\in\mathcal{A}^c})&=
h(\bbx|\{U_\ell\}_{\ell\in\mathcal{A}^c})-h(\bbx|U_1,\ldots,U_L) \\
&\geq h(\bbx|\{U_\ell\}_{\ell\in\mathcal{A}^c})-\frac{1}{2}
\log|(2\pi e)\mmse(\bbx|U_1,\ldots,U_L)|
\label{max_entropy}\\
&\geq h(\bbx|\{U_\ell\}_{\ell\in\mathcal{A}^c})-\frac{1}{2}
\log|(2\pi e)\bbd |
\label{distortion_constraint_implies}\\
&\geq h(\bbx|\{U_\ell\}_{\ell\in\mathcal{A}^c},W)-\frac{1}{2}
\log|(2\pi e)\bbd | \label{conditioning_reduces_entropy_y}
\end{align}
where (\ref{max_entropy}) comes from the fact that
$h(\bbx|U_1,\ldots,U_L)$ is maximized by jointly Gaussian
$(\bbx,U_1,\ldots,U_L)$, (\ref{distortion_constraint_implies})
follows from the monotonicity of $\log|\cdot|$ function in
positive semi-definite matrices in conjunction with the distortion
constraint in (\ref{distortion_constraint}), and
(\ref{conditioning_reduces_entropy_y}) comes from the fact that
conditioning cannot increase entropy.

Next, we obtain a lower bound for
$h(\bbx|\{U_\ell\}_{\ell\in\mathcal{A}^c},W)$. To this end, in
view of Lemma~\ref{lemma_dembo}, we note the following lower bound
on $h(\bbx|\{U_\ell\}_{\ell\in\mathcal{A}^c},W)$
\begin{align}
h(\bbx|\{U_\ell\}_{\ell\in\mathcal{A}^c},W)\geq \frac{1}{2}
\log|(2\pi e)\bbj^{-1}(\bbx|\{U_\ell\}_{\ell\in\mathcal{A}^c},W)|
\label{lower_bound_by_Fisher}
\end{align}
which implies that a lower bound on
$\bbj^{-1}(\bbx|\{U_\ell\}_{\ell\in\mathcal{A}^c},W)$ will yield a
lower bound for $h(\bbx|\{U_\ell\}_{\ell\in\mathcal{A}^c},W)$. To
obtain a lower bound for
$\bbj^{-1}(\bbx|\{U_\ell\}_{\ell\in\mathcal{A}^c},W)$, we will use
the connection between the Fisher information and the MMSE given
in Lemma~\ref{lemma_connection}. To this end, we note that $\bbx$
can be decomposed as (see (\ref{decomposition_reference}) in
Appendix~\ref{appendix_MMSE})
\begin{align}
\bbx &= \sum_{\ell\in\mathcal{A}^c} \bba_\ell
\bby_\ell+\bbn_{\mathcal{A}^c} \label{decomposition_of_X}
\end{align}
where the matrices $\{\bba_\ell\}_{\ell\in\mathcal{A}^c}$ are
given by (see (\ref{mmse_reference}) in
Appendix~\ref{appendix_MMSE})
\begin{align}
\bba_\ell=\bbsigma_{\mathcal{A}^c}\bbsigma_\ell^{-1},\quad
\ell\in\mathcal{A}^c \label{def_A_l}
\end{align}
In~(\ref{decomposition_of_X}), $\bbn_{\mathcal{A}^c}$ is a
zero-mean Gaussian vector with covariance matrix (see
(\ref{MMSE_matrix_reference_1}) in Appendix~\ref{appendix_MMSE})
\begin{align}
\bbsigma_{\mathcal{A}^c}=\left(\bbk_X^{-1}+\sum_{\ell\in\mathcal{A}^c}\bbsigma_\ell^{-1}\right)^{-1}
\label{def_estimation_error}
\end{align}
We also note that $\bbn_{\mathcal{A}^c}$ is independent of
$(\{\bby_\ell,U_{\ell}\}_{\ell\in\mathcal{A}^c},W)$ which implies
the following Markov chain
\begin{align}
\{U_{\ell}\}_{\ell\in\mathcal{A}^c},W \rightarrow
\sum_{\ell\in\mathcal{A}^c}\bba_\ell\bby_\ell\rightarrow
\bbx=\sum_{\ell\in\mathcal{A}^c}\bba_\ell\bby_\ell+\bbn_{\mathcal{A}^c}
\end{align}
In view of this Markov chain, due to Lemma~\ref{lemma_connection},
we have
\begin{align}
\mmse(\bbs_{\mathcal{A}^c}|\bbx,\{U_\ell\}_{\ell\in\mathcal{A}^c},W)=
\bbsigma_{\mathcal{A}^c}-\bbsigma_{\mathcal{A}^c}\bbj(\bbx|\{U_\ell\}_{\ell\in\mathcal{A}^c},W)\bbsigma_{\mathcal{A}^c}
\label{lemma_connection_implies}
\end{align}
where we define $\bbs_{\mathcal{A}^c}$ as follows
\begin{align}
\bbs_{\mathcal{A}^c}=\sum_{\ell\in\mathcal{A}^c}\bba_\ell\bby_\ell
\end{align}
Next, we obtain the MMSE matrix in
(\ref{lemma_connection_implies}) in terms of the individual MMSE
matrices
$\{\mmse(\bby_\ell|\bbx,U_\ell,W)\}_{\ell\in\mathcal{A}^c}$ as
given in the following lemma.
\begin{Lem}
\label{lemma_estimation} Under the current conditions, we have
\begin{align}
\mmse(\bbs_{\mathcal{A}^c}|\bbx,\{U_\ell\}_{\ell\in\mathcal{A}^c},W)&=
\sum_{\ell\in\mathcal{A}^c} \bba_\ell
\mmse(\bby_\ell|\bbx,W,U_\ell)\bba_\ell^\top
\end{align}
\end{Lem}
The proof of this lemma is given in
Appendix~\ref{proof_of_lemma_estimation}.

Hence, using Lemma~\ref{lemma_estimation} in
(\ref{lemma_connection_implies}), we get
\begin{align}
\bbsigma_{\mathcal{A}^c}-\bbsigma_{\mathcal{A}^c}\bbj(\bbx|\{U_\ell\}_{\ell\in\mathcal{A}^c},W)\bbsigma_{\mathcal{A}^c}&=
\sum_{\ell\in\mathcal{A}^c} \bba_\ell
\mmse(\bby_\ell|\bbx,W,U_\ell)\bba_\ell^\top \\
&\succeq \sum_{\ell\in\mathcal{A}^c} \bba_\ell
\bbd_\ell(\alpha_\ell^*)\bba_\ell^\top \label{order_1_implies}\\
&=\bbsigma_{\mathcal{A}^c}\left(\sum_{\ell\in\mathcal{A}^c}
\bbsigma_\ell^{-1}
\bbd_\ell(\alpha_\ell^*)\bbsigma_\ell^{-1}\right)\bbsigma_{\mathcal{A}^c}
\label{towards_a_lower_bound}
\end{align}
where (\ref{order_1_implies}) is due to (\ref{order_1}), and in
(\ref{towards_a_lower_bound}), we use the definition of
$\bba_\ell$ given in (\ref{def_A_l}). We note that
(\ref{towards_a_lower_bound}) implies
\begin{align}
\bbj^{-1}(\bbx|\{U_\ell\}_{\ell\in\mathcal{A}^c},W)&\succeq
\left(\bbsigma_{\mathcal{A}^c}^{-1}-\sum_{\ell\in\mathcal{A}^c}
\bbsigma_\ell^{-1}
\bbd_\ell(\alpha_\ell^*)\bbsigma_\ell^{-1}\right)^{-1}\\
&=\left(\bbk_X^{-1}+\sum_{\ell\in\mathcal{A}^c}\bbsigma_\ell^{-1}-\sum_{\ell\in\mathcal{A}^c}
\bbsigma_\ell^{-1}
\bbd_\ell(\alpha_\ell^*)\bbsigma_\ell^{-1}\right)^{-1}
\label{def_estimation_error_implies}
\end{align}
where (\ref{def_estimation_error_implies}) comes from the
definition of $\bbsigma_{\mathcal{A}^c}$ in
(\ref{def_estimation_error}). In view of
(\ref{lower_bound_by_Fisher}) and
(\ref{def_estimation_error_implies}), we have the following lower
bound for $h(\bbx|\{U_\ell\}_{\ell\in\mathcal{A}^c},W)$ as follows
\begin{align}
h(\bbx|\{U_\ell\}_{\ell\in\mathcal{A}^c},W)\geq \frac{1}{2}
\log\left|(2\pi e)
\left(\bbk_X^{-1}+\sum_{\ell\in\mathcal{A}^c}\bbsigma_\ell^{-1}-\sum_{\ell\in\mathcal{A}^c}
\bbsigma_\ell^{-1}
\bbd_\ell(\alpha_\ell^*)\bbsigma_\ell^{-1}\right)^{-1} \right|
\label{final_lower_bound}
\end{align}
Hence, using (\ref{final_lower_bound}) in
(\ref{conditioning_reduces_entropy_y}), we get
\begin{align}
I(\bbx;\{U_\ell\}_{\ell\in\mathcal{A}}|\{U_\ell\}_{\ell\in\mathcal{A}^c})\geq
\frac{1}{2} \log\frac{\left|
\left(\bbk_X^{-1}+\sum_{\ell\in\mathcal{A}^c}\bbsigma_\ell^{-1}-\sum_{\ell\in\mathcal{A}^c}
\bbsigma_\ell^{-1}
\bbd_\ell(\alpha_\ell^*)\bbsigma_\ell^{-1}\right)^{-1}
\right|}{|\bbd|}
\end{align}
Moreover, using the non-negativity of the mutual information, we
can improve this lower bound as follows
\begin{align}
I(\bbx;\{U_\ell\}_{\ell\in\mathcal{A}}|\{U_\ell\}_{\ell\in\mathcal{A}^c})\geq
\frac{1}{2} \log^+\frac{\left|
\left(\bbk_X^{-1}+\sum_{\ell\in\mathcal{A}^c}\bbsigma_\ell^{-1}-\sum_{\ell\in\mathcal{A}^c}
\bbsigma_\ell^{-1}
\bbd_\ell(\alpha_\ell^*)\bbsigma_\ell^{-1}\right)^{-1}
\right|}{|\bbd|} \label{half_of_the proof}
\end{align}
where $\log^+x=\max(\log x,0)$. Using (\ref{fixed_ones}) and
(\ref{half_of_the proof}) in the rate bounds given in
(\ref{outer_bound_general}), we get
\begin{align}
\sum_{\ell\in\mathcal{A}}R_\ell\geq \frac{1}{2} \log^+\frac{\left|
\left(\bbk_X^{-1}+\sum_{\ell\in\mathcal{A}^c}\bbsigma_\ell^{-1}-\sum_{\ell\in\mathcal{A}^c}
\bbsigma_\ell^{-1}
\bbd_\ell(\alpha_\ell^*)\bbsigma_\ell^{-1}\right)^{-1}
\right|}{|\bbd|}+\sum_{\ell\in\mathcal{A}}\frac{1}{2}
\log\frac{|\bbsigma_\ell|}{|\bbd_\ell(\alpha_\ell^*)|}
\label{rate_lower_bound_final}
\end{align}

Next, we establish a connection between $\bbd$ and
$(\bbd_1(\alpha_1^*),\ldots,\bbd_L(\alpha_L^*))$. To this end, by
taking $\mathcal{A}^c=\{1,\ldots,L\}$ in
(\ref{def_estimation_error_implies}), we get
\begin{align}
\left(\bbk_X^{-1}+\sum_{\ell=1}^L\bbsigma_\ell^{-1}-\sum_{\ell=1}^{L}
\bbsigma_\ell^{-1}
\bbd_\ell(\alpha_\ell^*)\bbsigma_\ell^{-1}\right)^{-1}&\preceq
\bbj^{-1}(\bbx|\{U_\ell\}_{\ell=1}^L,W) \\
&\preceq \mmse(\bbx|\{U_\ell\}_{\ell=1}^L,W) \label{lemma_conditional_crb_implies} \\
&\preceq \mmse(\bbx|\{U_\ell\}_{\ell=1}^L) \label{conditioning_reduces_MMSE_Y} \\
&\preceq \bbd \label{distortion_constraint_implies_1}
\end{align}
where (\ref{lemma_conditional_crb_implies}) is due to
Lemma~\ref{lemma_conditional_crb},
(\ref{conditioning_reduces_MMSE_Y}) comes from the fact that
conditioning reduces the MMSE matrix in the positive semi-definite
ordering sense, and (\ref{distortion_constraint_implies_1})
follows from the distortion constraint in
(\ref{distortion_constraint}). Hence, in view of
(\ref{rate_lower_bound_final}) and
(\ref{distortion_constraint_implies_1}), we show that the rate
region of the vector Gaussian CEO problem is included in the union
of rate tuples $(R_1,\ldots,R_L)$ satisfying
\begin{align}
\sum_{\ell\in\mathcal{A}}R_\ell\geq \frac{1}{2} \log^+\frac{\left|
\left(\bbk_X^{-1}+\sum_{\ell\in\mathcal{A}^c}\bbsigma_\ell^{-1}-\sum_{\ell\in\mathcal{A}^c}
\bbsigma_\ell^{-1}
\bbd_\ell(\alpha_\ell^*)\bbsigma_\ell^{-1}\right)^{-1}
\right|}{|\bbd|}+\sum_{\ell\in\mathcal{A}}\frac{1}{2}
\log\frac{|\bbsigma_\ell|}{|\bbd_\ell(\alpha_\ell^*)|}
\label{rate_lower_bound_final_again}
\end{align}
for all $\mathcal{A}\subseteq\{1,\ldots,L\}$, where the union is
over all positive semi-definite matrices
$\bbd_1(\alpha_1^*),\ldots,\break \bbd_L(\alpha_L^*)$ satisfying
the following orders
\begin{align}
\left(
\bbk_X^{-1}+\sum_{\ell=1}^L\bbsigma_\ell^{-1}-\sum_{\ell=1}^{L}
\bbsigma_\ell^{-1} \bbd_\ell(\alpha_\ell^*)\bbsigma_\ell^{-1}\right)^{-1} &\preceq \bbd\\
\bzero\preceq \bbd_{\ell}(\alpha_\ell^*) &\preceq
\bbsigma_\ell,\quad \ell=1,\ldots,L \label{additional_orders}
\end{align}
The orders in (\ref{additional_orders}) follow from
(\ref{less_than_noise}). The region given in
Theorem~\ref{theorem_outer} can be obtained from the outer bound
described in
(\ref{rate_lower_bound_final_again})-(\ref{additional_orders}) by
setting $\bbd_{\ell}(\alpha_\ell^*)=\bbd_\ell$, which completes
the proof of Theorem~\ref{theorem_outer}.

\section{Generalization of the Bounds}

In this section, we consider the most general form of the vector
Gaussian CEO problem, and generalize the outer and the inner
bounds in Theorem~\ref{theorem_outer} and
Theorem~\ref{theorem_inner}, respectively. In the most general
form of the vector Gaussian CEO problem, the observations at the
sensors are given by
\begin{align}
\bby_\ell=\bbh_\ell\bbx+\bbn_\ell,\quad \ell=1,\ldots,L
\label{general_form}
\end{align}
where $\{\bbn_\ell\}_{\ell=1}^L$ are i.i.d. zero-mean Gaussian
random vectors with identity covariance matrices. We note that the
general form for the observations in (\ref{general_form}) cover
the model in (\ref{aligned_observations}) we studied so far. All
definitions we introduced in Section~\ref{sec:aligned_model} hold
for the general model defined by (\ref{general_form}) except for
the distortion constraints in (\ref{limits_of_distortion}). In the
general model, the distortion $\bbd$ is assumed to satisfy
\begin{align}
\left(\bbk_X^{-1}+\sum_{\ell=1}^L
\bbh_\ell^\top\bbh_\ell\right)^{-1}\preceq \bbd \preceq \bbk_X
\label{limits_of_distortion_general}
\end{align}
where the left hand-side is the MMSE matrix obtained when the CEO
unit has access to all observations in (\ref{general_form}).
Similar to the model given by (\ref{aligned_observations}), here
also, imposing the lower bound constraint on $\bbd$ in
(\ref{limits_of_distortion_general}) does not incur any loss of
generality, while the upper bound constraint on $\bbd$ in
(\ref{limits_of_distortion_general}) might incur some loss of
generality.

Now, we provide an outer bound for the rate-distortion region
$\mathcal{R}(\bbd)$ for the general model given by
(\ref{general_form}), which, in fact, corresponds to the
generalization of the outer bound in Theorem~\ref{theorem_outer}
to the most general form of the vector Gaussian CEO problem.
\begin{Theo}
\label{theorem_outer_general} An outer bound for the
rate-distortion region of the general vector Gaussian CEO problem
is given by the union of rate tuples $(R_1,\ldots,R_L)$ satisfying
\begin{align}
\sum_{\ell\in\mathcal{A}}R_\ell&\geq \frac{1}{2}\log^+
\frac{\left|\left(\bbk_X^{-1}+\sum_{\ell\in\mathcal{A}^c}\bbh_\ell^\top
(\bbi-\bbd_\ell)\bbh_\ell\right)^{-1}\right|}{|\bbd|}
+\sum_{\ell\in\mathcal{A}}\frac{1}{2}\log\frac{1}{|\bbd_\ell|}
\end{align}
for all $\mathcal{A}\subseteq \{1,\ldots,L\}$, where the union is
over all positive semi-definite matrices
$\{\bbd_\ell\}_{\ell=1}^L$ satisfying the following constraints
\begin{align}
\left(\bbk_X^{-1}+\sum_{\ell=1}^L \bbh_\ell^\top(\bbi-\bbd_\ell)\bbh_\ell\right)^{-1} &\preceq \bbd
\label{distortion_constraint_general}\\
\bzero \preceq \bbd_\ell & \preceq \bbi,\quad \ell=1,\ldots,L
\end{align}
\end{Theo}

We prove Theorem~\ref{theorem_outer_general} in two steps. In the
first step, we enhance (improve) the observations at the sensors
in a way that the enhanced observations are in a similar form
given by (\ref{aligned_observations}). In the next step, we use
Theorem~\ref{theorem_outer} to obtain an outer bound for the
enhanced model, and from this outer bound, we obtain
Theorem~\ref{theorem_outer_general} by using some limiting
arguments. The details of the proof can be found in
Appendix~\ref{proof_of_theorem_outer_general}.

Now, we introduce an inner bound for the rate-distortion region
$\mathcal{R}(\bbd)$ for the general model given by
(\ref{general_form}), which, in fact, corresponds to the
generalization of the inner bound in Theorem~\ref{theorem_inner}
to the most general form of the vector Gaussian CEO problem.
\begin{Theo}
\label{theorem_inner_general} An inner bound for the
rate-distortion region of the general vector Gaussian CEO problem
is given by the union of rate tuples $(R_1,\ldots,R_L)$ satisfying
\begin{align}
\sum_{\ell\in\mathcal{A}}R_\ell \geq
\frac{1}{2}\log\frac{\left|\left(\bbk_X^{-1}+\sum_{\ell\in\mathcal{A}^c}\bbh_\ell^\top
(\bbi-\bbd_\ell)\bbh_\ell\right)^{-1}\right|}
{\left|\left(\bbk_X^{-1}+\sum_{\ell=1}^L \bbh_\ell^\top
(\bbi-\bbd_\ell)\bbh_\ell\right)^{-1}\right|}
+\sum_{\ell\in\mathcal{A}}\frac{1}{2} \log\frac{1}{|\bbd_\ell|}
\label{rate_upper_bound_Gaussian_final_general}
\end{align}
for all $\mathcal{A}\subseteq \{1,\ldots,L\}$, where the union is
over all positive semi-definite matrices
$\{\bbd_\ell\}_{\ell=1}^L$ satisfying
\begin{align}
\left(\bbk_X^{-1}+\sum_{\ell=1}^L \bbh_\ell^\top
(\bbi-\bbd_\ell)\bbh_\ell\right)^{-1}&\preceq
\bbd \label{dummy_identity_implies_implies_general}\\
\bzero \preceq \bbd_\ell &\preceq \bbi,\quad \ell=1,\ldots,L
\label{orders_of_D_ell_implies_general}
\end{align}
\end{Theo}

The proof of Theorem~\ref{theorem_inner_general} is given in
Appendix~\ref{sec:proof_of_inner_bound}. We obtain this inner
bound by evaluating the Berger-Tung inner bound~\cite{thesis_tung}
by jointly Gaussian auxiliary random variables.

We note that since the outer and the inner bounds in
Theorem~\ref{theorem_outer_general} and
Theorem~\ref{theorem_inner_general} correspond to the
generalizations of the outer and inner bounds in
Theorem~\ref{theorem_outer} and Theorem~\ref{theorem_inner},
respectively, our previous comments and remarks about
Theorem~\ref{theorem_outer} and Theorem~\ref{theorem_inner} hold
for Theorem~\ref{theorem_outer_general} and
Theorem~\ref{theorem_inner_general} as well. In particular,
similar to Theorem~\ref{theorem_outer} and
Theorem~\ref{theorem_inner}, we can provide alternative
characterizations for Theorem~\ref{theorem_outer_general} and
Theorem~\ref{theorem_inner_general} as well. Moreover, similar to
Theorem~\ref{theorem_outer} and Theorem~\ref{theorem_inner}, the
bounds in Theorem~\ref{theorem_outer_general} and
Theorem~\ref{theorem_inner_general} match when the boundary of the
outer bound in Theorem~\ref{theorem_outer_general} can be
described by the matrices $\{\bbd_\ell\}_{\ell=1}^L$ that satisfy
the distortion constraint in (\ref{distortion_constraint_general})
with equality.

\section{Conclusions}
In this paper, we study the vector Gaussian CEO problem and
provide an outer bound for its rate-distortion region. We obtain
our outer bound by evaluating the rather general outer bound
in~\cite{Wagner_Outer_Bound}. We accomplish this evaluation by
using a technique that relies on the de Bruijn identity along with
the properties of the MMSE and Fisher information. We show that
our outer bound strictly improves the existing outer bounds by
providing an example, in which, our outer bound is strictly
contained in the existing outer bounds. However, despite this
improvement, we show that our outer bound does not provide the
exact rate-distortion region in general. We show this by providing
an example where our outer bound strictly includes the
rate-distortion region.

\appendixpage
\appendices

\section{Distortion Limits}
\label{appendix_distortion_limits}

In this appendix, we first note some facts about Gaussian random
vectors that are used throughout the paper.

\subsection{Gaussian Random Vectors}
\label{appendix_MMSE} Let $\bbt$ be a zero-mean Gaussian random
vector with covariance matrix $\bbsigma_T\succ\bzero$. We define
the Gaussian random vectors $\{\bbt_\ell\}_{\ell=1}^{L}$ as
\begin{align}
\bbt_\ell&=\bbh_\ell\bbt+\bbn_\ell
\end{align}
where $\{\bbn_\ell\}_{\ell=1}^L$ are zero-mean independent
Gaussian random vectors with covariance matrices
$\{\bbsigma_\ell\}_{\ell=1}^L$, which are also independent of
$\bbt$. We assume $\bbsigma_{\ell}\succ \bzero,~\ell=1,\ldots,L$.

For any subset $\mathcal{A}\subseteq \{1,\ldots,L\}$, we have
\begin{align}
\bbt&=\sum_{\ell\in\mathcal{A}}\bba_\ell \bbt_\ell
+\bbn_{\mathcal{A}} \label{decomposition_reference}
\end{align}
where $\bbn_{\mathcal{A}}$ is a zero-mean Gaussian random vector
with covariance matrix $\bbsigma_{\mathcal{A}}$ given by
\begin{align}
\bbsigma_{\mathcal{A}}=\left(\bbsigma_T^{-1}+\sum_{\ell\in\mathcal{A}}\bbh_\ell^\top\bbsigma_\ell^{-1}\bbh_\ell\right)^{-1}
\label{MMSE_matrix_reference_1}
\end{align}
and is independent of $\{\bbt_\ell\}_{\ell\in\mathcal{A}}$. The
matrices $\{\bba_\ell\}_{\ell\in\mathcal{A}}$ are given by
\begin{align}
\bba_\ell=\bbsigma_{\mathcal{A}}\bbh_\ell^\top\bbsigma_\ell^{-1},\quad
\ell\in\mathcal{A} \label{mmse_reference}
\end{align}

The decomposition in (\ref{decomposition_reference}) follows from
the MMSE estimation of Gaussian random vectors, which is
equivalent to the linear MMSE estimation. In particular, we have
\begin{align}
\hat{\bbt}&=E\left[\bbt|\{\bbt_\ell\}_{\ell\in\mathcal{A}}\right]=\sum_{\ell\in\mathcal{A}}\bba_\ell\bbt_\ell
\end{align}
which is the MMSE, equivalently the linear MMSE, estimator of
$\bbt$ from $\{\bbt_\ell\}_{\ell=1}^L$. The error in estimation is
$\bbn_\mathcal{A}$, and the MMSE matrix is
\begin{align}
\mmse(\bbt|\{\bbt_\ell\}_{\ell\in\mathcal{A}})=\bbsigma_{\mathcal{A}}
\label{MMSE_matrix_reference_2}
\end{align}

\subsection{Regarding (\ref{limits_of_distortion})}

We first obtain the lower bound on the distortion constraint
$\bbd$ in (\ref{limits_of_distortion}) as follows
\begin{align}
\mmse(\bbx_i|B_1^n,\ldots,B_L^n)
&\succeq \mmse(\bbx_i|B_1^n,\ldots,B_L^n,\bby_1^n,\ldots,\bby_L^n)\label{conditioning_reduces_entropy_x}\\
&=\mmse(\bbx_i|\bby_1^n,\ldots,\bby_L^n) \label{functions}\\
&=\mmse(\bbx_i|\bby_{1,i},\ldots,\bby_{L,i}) \label{independence_across_time}\\
&=\left(\bbk_X^{-1}+\sum_{\ell=1}^L \bbsigma_\ell^{-1}\right)^{-1}
\label{mmse_reference_implies}
\end{align}
where (\ref{conditioning_reduces_entropy_x}) follows from the fact
that conditioning reduces the MMSE matrix in the positive
semi-definite ordering sense, (\ref{functions}) is due to the fact
that $B_\ell^n$ is a function of $\bby_\ell^n$,
(\ref{independence_across_time}) comes from the independence of
$(\bbx_i,\bby_{1,i},\ldots,\bby_{L,i})$ across time, and
(\ref{mmse_reference_implies}) is due to
(\ref{MMSE_matrix_reference_1}) and
(\ref{MMSE_matrix_reference_2}). Hence,
(\ref{mmse_reference_implies}) implies that imposing the
constraint $\bbd\succeq \left(\bbk_X^{-1}+\sum_{\ell=1}^L
\bbsigma_\ell^{-1}\right)^{-1} $ does not incur any loss of
generality.

Next, we consider the upper bound on the distortion constraint in
(\ref{distortion_constraint}). To this end, we note the following
order
\begin{align}
\mmse(\bbx_i|B_1^n,\ldots,B_L^n)\preceq \mmse(\bbx)=\bbk_X
\label{always_smaller_than_noise}
\end{align}
where we use the fact that conditioning reduces the MMSE matrix in
the positive semi-definite ordering sense. Equation
(\ref{always_smaller_than_noise}) implies that all
$(n,R_1,\ldots,R_L)$ codes achieve a distortion which is smaller
than $\bbk_X$. In other words, if $\hat{\bbd}$ is the distortion
achieved by a specific code, we always have $\hat{\bbd}\preceq
\bbk_X$. In spite of this fact, we still cannot impose the
constraint $\bbd\preceq \bbk_X$ without loss of generality. To
demonstrate this point, assume that $\bbk_X-\bbd$ is indefinite.
Hence, to be able to impose the constraint $\bbd\preceq \bbk_X$,
we should find a new distortion constraint $ \bbd^\prime$ which
satisfies $\bbd^{\prime}\preceq\{\bbd,\bbk_X\}$ and the
rate-distortion regions $\mathcal{R}(\bbd)$ and
$\mathcal{R}(\bbd^{\prime})$ are identical. In other words, there
needs to be a distortion matrix $\bbd^\prime\preceq
\{\bbd,\bbk_X\}$, and for any code achieving a distortion
$\hat{\bbd}\preceq \bbd$, we also have $\hat{\bbd}\preceq
\bbd^\prime$. However, as we will show now, this is not possible
in general. Assume that there exist two codes achieving the
distortion $\hat{\bbd}_j,~j=1,2,$ where $\hat{\bbd}_j\preceq
\{\bbd,\bbk_X\}$. Hence, $\bbd^\prime$ needs to satisfy
\begin{align}
\{\hat{\bbd}_1,\hat{\bbd}_2\}\preceq \bbd^{\prime} \preceq
\{\bbd,\bbk_X\} \label{nothing_between}
\end{align}
However, there are cases where it is impossible to find a matrix
$\bbd^\prime$ satisfying the order in (\ref{nothing_between}) as
shown in~\cite[Appendix~I]{Liu_Compound} by a counter-example.
Consequently, imposing the constraint $\bbd\preceq \bbk_X$ might
incur some loss of generality.

\section{Proofs of Theorem~\ref{theorem_outer_tangent} and Theorem~\ref{theorem_inner_tangent}}
\label{proof_of_tangent_line_bounds}

Here, we prove only Theorem~\ref{theorem_outer_tangent}. The proof
of Theorem~\ref{theorem_inner_tangent} is similar to the proof of
Theorem~\ref{theorem_outer_tangent}, and can be concluded from the
proof we present here. First, we note that our outer bound in
Theorem~\ref{theorem_outer} can be expressed as
\begin{align}
\sum_{\ell\in\mathcal{A}}\bar{R}_\ell \geq f(\mathcal{A}),\quad
\mathcal{A}\subseteq \{1,\ldots,L\}
\label{outer_bound_contra_polymatroid}
\end{align}
where
\begin{align}
\bar{R}_\ell&=R_\ell-\frac{1}{2}
\log\frac{|\bbsigma_\ell|}{|\bbd_\ell|},\quad \ell=1,\ldots,L
\label{alternative_rates} \\
f(\mathcal{A})&=\frac{1}{2}\log^+\frac{\left|\left(\bbk_X^{-1}+\sum_{\ell\in\mathcal{A}^c}\bbsigma_\ell^{-1}\left(
\bbsigma_\ell-\bbd_\ell\right)\bbsigma_\ell^{-1}\right)^{-1}\right|}{|\bbd|},\quad
\forall \mathcal{A}\subseteq \{1,\ldots,L\} \label{set_function}
\end{align}
Next, we show that $f(\mathcal{A})$ satisfies the following
properties.
\begin{Lem}
\label{Lemma_contra_polymatroid}
\begin{align}
f(\emptyset)&=0 \label{set_function_empty_set}\\
f(\mathcal{A }\cup \{t\})&\geq f(\mathcal{A}),\quad \forall
t\in\{1,\ldots,L\} \label{set_function_monotonicity}\\
f(\mathcal{A}\cup\mathcal{B})+f(\mathcal{A}\cap\mathcal{B})&\geq
f(\mathcal{A})+f(\mathcal{B}) \label{set_function_super_modular}
\end{align}
\end{Lem}
The proof of Lemma~\ref{Lemma_contra_polymatroid} is given in
Appendix~\ref{proof_of_Lemma_contra_polymatroid}.

A set function $f(\mathcal{A})$ satisfying the properties in
Lemma~\ref{Lemma_contra_polymatroid} is called a supermodular
function. The region defined by means of a supermodular function
as in~(\ref{outer_bound_contra_polymatroid}) is called a
contra-polymatroid~\cite{Welsh_Matroid}. We denote the
contra-polymatroid defined in
(\ref{outer_bound_contra_polymatroid}) by $\mathcal{G}(f)$. An
important property of contra-polymatroids is that all of their
vertices can be found in an explicit form. In particular, when
$\mu_1\geq \ldots\mu_L\geq 0$, the vertex corresponding to the
tangent hyperplane $\sum_{\ell=1}^L \mu_\ell \bar{R}_\ell$ is
given by~\cite[Lemma 3.3]{Tse_MAC_Fading}
\begin{align}
\bar{R}_\ell^*=f(\{1,\ldots,\ell\})-f(\{1,\ldots,\ell-1\}),\quad
\ell=1,\ldots,L \label{vertex}
\end{align}
using which, we have
\begin{align}
\lefteqn{\min_{(\bar{R}_1,\ldots,\bar{R}_L)\in\mathcal{G}(f)}\sum_{\ell=1}^L
\mu_\ell \bar{R}_\ell=\sum_{\ell=1}^L
\mu_\ell\big(f(\{1,\ldots,\ell\})-f(\{1,\ldots,\ell-1\})\big)} \\
&=\sum_{\ell=1}^L \mu_\ell
f(\{1,\ldots,\ell\})-\sum_{\ell=1}^{L-1}\mu_{\ell+1}
f(\{1,\ldots,\ell\})\\
&=\sum_{\ell=1}^{L-1}\frac{\mu_\ell-\mu_{\ell+1}}{2}\log^+\frac{\left|\left(\bbk_X^{-1}+\sum_{j=\ell+1}^L
\bbsigma_j^{-1}\left(
\bbsigma_j-\bbd_j\right)\bbsigma_j^{-1}\right)^{-1}\right|}{|\bbd|}
+\frac{\mu_L}{2}  \log\frac{|\bbk_X|}{|\bbd|}
\end{align}
which, in turn, implies
\begin{align}
\min_{(R_1,\ldots,R_L)\in\mathcal{R}(\bbd)}~\sum_{\ell=1}^L\mu_\ell
R_\ell&=\min_{\{\bbd_\ell\}_{\ell=1}^L}~\sum_{\ell=1}^{L-1}\frac{\mu_\ell-\mu_{\ell+1}}{2}\log^+\frac{\left|\left(\bbk_X^{-1}+\sum_{j=\ell+1}^L
\bbsigma_j^{-1}\left(
\bbsigma_j-\bbd_j\right)\bbsigma_j^{-1}\right)^{-1}\right|}{|\bbd|}\nonumber\\
&\qquad \qquad \quad
+\sum_{\ell=1}^L\frac{\mu_\ell}{2}\log\frac{|\bbsigma_\ell|}{|\bbd_\ell|}
+\frac{\mu_L}{2} \log\frac{|\bbk_X|}{|\bbd|}
\label{final_alternative}
\end{align}
where $\{\bbd_\ell\}_{\ell=1}^L$ are subject to the constraints in
(\ref{feasible_set_outer_I_tangent})-(\ref{feasible_set_outer_II_tangent}).
Since (\ref{final_alternative}) is the desired result
in~Theorem~\ref{theorem_outer_tangent}; this completes the proof
of Theorem~\ref{theorem_outer_tangent}.

\subsection{Proof of Lemma~\ref{Lemma_contra_polymatroid}}

\label{proof_of_Lemma_contra_polymatroid}

Now we prove Lemma~\ref{Lemma_contra_polymatroid}. The first
property of the set function $f(\mathcal{A})$ given in
(\ref{set_function_empty_set}) is immediate by noting from
(\ref{feasible_set_outer_I}) that
\begin{align}
\left(\bbk_X^{-1}+\sum_{\ell=1}^L
\bbsigma_\ell^{-1}-\bbsigma_{\ell}^{-1}\bbd_\ell\bbsigma_\ell^{-1}\right)^{-1}\preceq
\bbd
\end{align}

Next, we prove (\ref{set_function_monotonicity}) and
(\ref{set_function_super_modular}). To this end, we define the
jointly Gaussian random vector tuple $(U_1^*,\ldots,U_L^*)$ which
satisfies the Markov chain
\begin{align}
U_i^*\rightarrow \bbx \rightarrow \{U_j^*\}_{j=1,j\neq i}^L,\quad
i=1,\ldots,L \label{MC_inclusion}
\end{align}
and
\begin{align}
h(\bbx|\{U_\ell^*\}_{\ell\in\mathcal{A}^c})=\frac{1}{2} \log\left|
\left(\bbk_X^{-1}+\sum_{\ell\in\mathcal{A}^c}\bbsigma_\ell^{-1}-\sum_{\ell\in\mathcal{A}^c}
\bbsigma_\ell^{-1} \bbd_\ell\bbsigma_\ell^{-1}\right)^{-1} \right|
\label{conditional_entropy}
\end{align}
for all $\mathcal{A}^c\subseteq \{1,\ldots,L\}$. Here, we do not
show the existence of the jointly Gaussian random vector tuple
$(U_1^*,\ldots,U_L^*)$ satisfying
(\ref{MC_inclusion})-(\ref{conditional_entropy}), however the
existence of such Gaussian random vector tuples can be concluded
from the analysis in Appendix~\ref{sec:proof_of_inner_bound} where
we prove Theorem~\ref{theorem_inner} (the inner bound for the
rate-distortion region). Hence, using $(U_1^*,\ldots,U_L^*)$, the
set function $f(\mathcal{A})$ can be written as
\begin{align}
f(\mathcal{A})=\max\left(0,h(\bbx|\{U^*_\ell\}_{\ell\in\mathcal{A}^c})-\frac{1}{2}\log|(2\pi
e)\bbd|\right),\quad \mathcal{A}\subseteq\{1,\ldots,L\}
\end{align}
The monotonicity of the set function $f(\mathcal{A})$ can be shown
as follows
\begin{align}
f(\mathcal{A}\cup\{t\})&=\max\left(0,h(\bbx|\{U^*_\ell\}_{\ell\in\mathcal{A}^c,\ell\neq
t})-\frac{1}{2}\log|(2\pi e)\bbd|\right)\\
&\geq
\max\left(0,h(\bbx|\{U^*_\ell\}_{\ell\in\mathcal{A}^c})-\frac{1}{2}\log|(2\pi
e)\bbd|\right) \label{conditioning_reduces_q_1}\\
&=f(\mathcal{A}) \label{monotonicity_is_proved}
\end{align}
where (\ref{conditioning_reduces_q_1}) follows from the fact that
conditioning cannot increase entropy. Equation
(\ref{monotonicity_is_proved}) proves
(\ref{set_function_monotonicity}).

Finally, we consider (\ref{set_function_super_modular}) as follows
\begin{align}
\lefteqn{\hspace{-0.5cm}f(\mathcal{A}\cup\mathcal{B})+f(\mathcal{A}\cap\mathcal{B})}\nonumber\\
&=\max\left(0,h(\bbx|\{U^*_\ell\}_{\ell\in\mathcal{A}^c\cap\mathcal{B}^c})-\frac{\alpha}{2}\right)+
\max\left(0,h(\bbx|\{U^*_\ell\}_{\ell\in\mathcal{A}^c\cup\mathcal{B}^c})-\frac{\alpha}{2}\right) \\
&=\max\left(0,h(\bbx|\{U^*_\ell\}_{\ell\in\mathcal{A}^c\cap\mathcal{B}^c})-\frac{\alpha}{2},h(\bbx|\{U^*_\ell\}_{\ell\in\mathcal{A}^c\cap\mathcal{B}^c})+
h(\bbx|\{U^*_\ell\}_{\ell\in\mathcal{A}^c\cup\mathcal{B}^c})-\alpha
\right)\label{inclusion_x_1}\\
&\geq\max\left(0,h(\bbx|\{U^*_\ell\}_{\ell\in\mathcal{A}^c})-\frac{\alpha}{2},h(\bbx|\{U^*_\ell\}_{\ell\in\mathcal{B}^c})-\frac{\alpha}{2},\right.\nonumber\\
&\qquad \qquad \quad
h(\bbx|\{U^*_\ell\}_{\ell\in\mathcal{A}^c\cap\mathcal{B}^c})+
h(\bbx|\{U^*_\ell\}_{\ell\in\mathcal{A}^c\cup\mathcal{B}^c})-\alpha
 \Big)\label{inclusion_x_2}
\end{align}
where $\alpha=\log|(2\pi e)\bbd|$, and
(\ref{inclusion_x_1})-(\ref{inclusion_x_2}) follow from
\begin{align}
h(\bbx|\{U^*_\ell\}_{\ell\in\mathcal{A}^c\cap\mathcal{B}^c})&\geq
h(\bbx|\{U^*_\ell\}_{\ell\in\mathcal{A}^c\cup\mathcal{B}^c})\\
h(\bbx|\{U^*_\ell\}_{\ell\in\mathcal{A}^c\cap\mathcal{B}^c})&\geq
\max\big(h(\bbx|\{U^*_\ell\}_{\ell\in\mathcal{A}^c}),h(\bbx|\{U^*_\ell\}_{\ell\in\mathcal{B}^c})\big)
\end{align}
respectively, which, in turn, come from the fact that conditioning
cannot increase entropy. Next, we consider the last term in
(\ref{inclusion_x_2}) as follows
\begin{align}
\lefteqn{h(\bbx|\{U^*_\ell\}_{\ell\in\mathcal{A}^c\cap\mathcal{B}^c})+
h(\bbx|\{U^*_\ell\}_{\ell\in\mathcal{A}^c\cup\mathcal{B}^c})}\nonumber\\
& =h(\bbx,\{U^*_\ell\}_{\ell\in\mathcal{A}^c\cap\mathcal{B}^c})+
h(\bbx,\{U^*_\ell\}_{\ell\in\mathcal{A}^c\cup\mathcal{B}^c})-h(\{U^*_\ell\}_{\ell\in\mathcal{A}^c\cap\mathcal{B}^c})-
h(\{U^*_\ell\}_{\ell\in\mathcal{A}^c\cup\mathcal{B}^c})\\
&=2h(\bbx)+\sum_{\ell\in\mathcal{A}^c}h(U_\ell^*|\bbx)+\sum_{\ell\in\mathcal{B}^c}h(U_\ell^*|\bbx)
-h(\{U^*_\ell\}_{\ell\in\mathcal{A}^c\cap\mathcal{B}^c})-
h(\{U^*_\ell\}_{\ell\in\mathcal{A}^c\cup\mathcal{B}^c})
\label{MC_inclusion_implies_x_1}\\
&=2h(\bbx)+\sum_{\ell\in\mathcal{A}^c}h(U_\ell^*|\bbx)+\sum_{\ell\in\mathcal{B}^c}h(U_\ell^*|\bbx)
-h(\{U^*_\ell\}_{\ell\in\mathcal{A}^c\cap\mathcal{B}^c})-h(\{U^*_\ell\}_{\ell\in\mathcal{A}^c})\nonumber\\
&\quad -
h(\{U^*_\ell\}_{\ell\in\mathcal{A}\cap\mathcal{B}^c}|\{U_\ell^*\}_{\ell\in\mathcal{A}^c})
\\
&=h(\bbx|\{U_\ell^*\}_{\ell\in\mathcal{A}^c})+h(\bbx)+\sum_{\ell\in\mathcal{B}^c}h(U_\ell^*|\bbx)
-h(\{U^*_\ell\}_{\ell\in\mathcal{A}^c\cap\mathcal{B}^c}) -
h(\{U^*_\ell\}_{\ell\in\mathcal{A}\cap\mathcal{B}^c}|\{U_\ell^*\}_{\ell\in\mathcal{A}^c})
\\
&\geq
h(\bbx|\{U_\ell^*\}_{\ell\in\mathcal{A}^c})+h(\bbx)+\sum_{\ell\in\mathcal{B}^c}h(U_\ell^*|\bbx)
-h(\{U^*_\ell\}_{\ell\in\mathcal{A}^c\cap\mathcal{B}^c}) -
h(\{U^*_\ell\}_{\ell\in\mathcal{A}\cap\mathcal{B}^c}|\{U_\ell^*\}_{\ell\in\mathcal{A}^c\cap\mathcal{B}^c}) \label{conditioning_reduces_q_2}\\
&=h(\bbx|\{U_\ell^*\}_{\ell\in\mathcal{A}^c})+h(\bbx)+\sum_{\ell\in\mathcal{B}^c}h(U_\ell^*|\bbx)
-h(\{U^*_\ell\}_{\ell\in\mathcal{B}^c}) \\
&=h(\bbx|\{U_\ell^*\}_{\ell\in\mathcal{A}^c})+
h(\bbx|\{U_\ell^*\}_{\ell\in\mathcal{B}^c}) \label{dummy_step}
\end{align}
where (\ref{MC_inclusion_implies_x_1}) comes from the Markov chain
in (\ref{MC_inclusion}), and (\ref{conditioning_reduces_q_2})
follows from the fact that conditioning cannot increase entropy.
Using (\ref{dummy_step}) in (\ref{inclusion_x_2}), we get
\begin{align}
\lefteqn{\!\!\!\! \!\!\! f(\mathcal{A}\cup\mathcal{B})+f(\mathcal{A}\cap\mathcal{B})}\nonumber\\
&\geq
\max\left(0,h(\bbx|\{U^*_\ell\}_{\ell\in\mathcal{A}^c})-\frac{\alpha}{2},h(\bbx|\{U^*_\ell\}_{\ell\in\mathcal{B}^c})-\frac{\alpha}{2},\right.\nonumber\\
&\qquad \qquad \quad h(\bbx|\{U^*_\ell\}_{\ell\in\mathcal{A}^c})+
h(\bbx|\{U^*_\ell\}_{\ell\in\mathcal{B}^c})-\alpha \Big)\\
&=\max\left(0,h(\bbx|\{U^*_\ell\}_{\ell\in\mathcal{A}^c})-\frac{\alpha}{2}\right)
+\max\left(0,h(\bbx|\{U^*_\ell\}_{\ell\in\mathcal{B}^c})-\frac{\alpha}{2}\right)\\
&=f(\mathcal{A})+f(\mathcal{B})
\end{align}
which proves (\ref{set_function_super_modular}); completing the
proof of Lemma~\ref{Lemma_contra_polymatroid}.

\section{Proofs of Corollaries~\ref{Corollary_after_assumptions} and~\ref{Corollary_after_assumptions_chen}}

\subsection{Proof of Corollary~\ref{Corollary_after_assumptions}}
\label{sec:proof_of_corollary_after_assumptions}

We define the set $\mathcal{D}^{++}(\bbd_1,\bbd_2)$ as the union
of $(\bbd_1,\bbd_2)$ satisfying
\begin{align}
\left(\bbk_X^{-1}+\sum_{\ell=1}^2\bbsigma_\ell^{-1}-\sum_{\ell=1}^2\bbsigma_\ell^{-1}\bbd_\ell\bbsigma_\ell^{-1}\right)^{-1}&\preceq
\bbd \label{new_feasible_set_I} \\
\bzero\preceq \bbd_1 &\preceq \bbsigma_1
\label{new_feasible_set_II}
\end{align}
We note that $\mathcal{D}^{+}(\bbd_1,\bbd_2)\subseteq
\mathcal{D}^{++}(\bbd_1,\bbd_2)$. Using this in
Corollary~\ref{corollary_our_outer_bound}, we have
\begin{align}
{\rm T^+}&\geq
\min_{(\bbd_1,\bbd_2)\in\mathcal{D}^{++}(\bbd_1,\bbd_2)}~\frac{\mu_1}{2}
\log\frac{|\bbsigma_1|}{|\bbd_1|}+\frac{\mu_2}{2}\log\frac{|\bbsigma_2|}{|\bbd_2|}
+\frac{\mu_2}{2}\log\frac{|\bbk_X|}{|\bbd|}\nonumber \\
&\qquad \qquad \qquad \qquad \qquad + \frac{\mu_1-\mu_2}{2}
\log^{+}\frac{\left|\left(\bbk_X^{-1}+\bbsigma_2^{-1}-\bbsigma_2^{-1}\bbd_2\bbsigma_2^{-1}\right)^{-1}\right|}{|\bbd|}
\\
&\geq
\min_{(\bbd_1,\bbd_2)\in\mathcal{D}^{++}(\bbd_1,\bbd_2)}~\frac{\mu_1}{2}
\log\frac{|\bbsigma_1|}{|\bbd_1|}+\frac{\mu_2}{2}\log\frac{|\bbsigma_2|}{|\bbd_2|}
+\frac{\mu_2}{2}\log\frac{|\bbk_X|}{|\bbd|}
\\
&=
\min_{(\bbd_1,\bbd_2)\in\mathcal{D}^{++}(\bbd_1,\bbd_2)}~\frac{\mu_1}{2}
\log\frac{|\bbsigma_1^{-1}|}{|\bbsigma_1^{-1}\bbd_1\bbsigma_1^{-1}|}+\frac{\mu_2}{2}\log\frac{|\bbsigma_2^{-1}|}{|\bbsigma_2^{-1}\bbd_2\bbsigma_2^{-1}|}
+\frac{\mu_2}{2}\log\frac{|\bbk_X|}{|\bbd|}
\\
&\geq \min_{\bzero \preceq \bbd_1\preceq
\bbsigma_1}~\frac{\mu_1}{2}
\log\frac{|\bbsigma_1^{-1}|}{|\bbsigma_1^{-1}\bbd_1\bbsigma_1^{-1}|}
+\frac{\mu_2}{2}\log\frac{|\bbsigma_2^{-1}|}{|\bbk_X^{-1}+\bbsigma_1^{-1}+\bbsigma_2^{-1}-\bbsigma_1^{-1}\bbd_1\bbsigma_1^{-1}-\bbd^{-1}|}\nonumber\\
&\qquad \qquad \qquad +\frac{\mu_2}{2}\log\frac{|\bbk_X|}{|\bbd|}
\label{monotonicity_implies}
\end{align}
where we obtain (\ref{monotonicity_implies}) by using the fact
that $\log|\bbsigma_2^{-1}\bbd_2\bbsigma_2^{-1}|$ is monotonically
increasing in positive semi-definite matrices $\bbd_2$ and the
order on $\bbd_2$ given in (\ref{new_feasible_set_I}). Next, we
show that the cost function in (\ref{monotonicity_implies}) is
monotonically decreasing in $\bbd_1$, or equivalently in
$\bbsigma_1^{-1}\bbd_1\bbsigma_1^{-1}$. To this end, we consider
the gradient of the cost function in (\ref{monotonicity_implies})
with respect to the matrix $\bbsigma_1^{-1}\bbd_1\bbsigma_1^{-1}$,
which is equivalent to
\begin{align}
-\mu_1
\left(\bbsigma_1^{-1}\bbd_1\bbsigma_1^{-1}\right)^{-1}+\mu_2
\left(\bbk_X^{-1}+\bbsigma_1^{-1}+\bbsigma_2^{-1}-\bbsigma_1^{-1}\bbd_1\bbsigma_1^{-1}-\bbd^{-1}\right)^{-1}
\label{gradient}
\end{align}
Next, we show that (\ref{gradient}) is strictly negative definite;
implying that the cost function in (\ref{monotonicity_implies}) is
monotonically decreasing in $\bbd_1$. To this end, using the
assumption in (\ref{first_assumption}), we have
\begin{align}
\frac{1}{\mu_2} \left(
\bbk_X^{-1}+\bbsigma_1^{-1}+\bbsigma_2^{-1}-\bbd^{-1}\right)&\succ
\left(\frac{1}{\mu_1}+\frac{1}{\mu_2}\right)\bbsigma_1^{-1}\\
&\succeq
\left(\frac{1}{\mu_1}+\frac{1}{\mu_2}\right)\bbsigma_1^{-1}\bbd_1\bbsigma_1^{-1}
\label{smaller_implies}
\end{align}
where we use the fact that $ \bbd_1\preceq \bbsigma_1$. We note
that the order in (\ref{smaller_implies}) can be written as
\begin{align}
\frac{1}{\mu_2} \left(
\bbk_X^{-1}+\bbsigma_1^{-1}+\bbsigma_2^{-1}-\bbsigma_1^{-1}\bbd_1\bbsigma_1^{-1}-\bbd^{-1}\right)&\succ
\frac{1}{\mu_1}\bbsigma_1^{-1}\bbd_1\bbsigma_1^{-1}
\label{smaller_implies_1}
\end{align}
which is equivalent to
\begin{align}
\mu_2 \left(
\bbk_X^{-1}+\bbsigma_1^{-1}+\bbsigma_2^{-1}-\bbsigma_1^{-1}\bbd_1\bbsigma_1^{-1}-\bbd^{-1}\right)^{-1}&\prec
\mu_1\left(\bbsigma_1^{-1}\bbd_1\bbsigma_1^{-1}\right)^{-1}
\label{smaller_implies_2}
\end{align}
which, in turn, implies that the gradient of the cost function in
(\ref{monotonicity_implies}) is negative definite, and hence, the
cost function in (\ref{monotonicity_implies}) is monotonically
decreasing in $\bbd_1$. Consequently, this implies that the
minimum in (\ref{monotonicity_implies}) is attained when
$\bbd_1=\bbsigma_1$, i.e., we have
\begin{align}
{\rm T^+} &\geq
\frac{\mu_2}{2}\log\frac{|\bbsigma_2^{-1}|}{|\bbk_X^{-1}+\bbsigma_2^{-1}-\bbd^{-1}|}+\frac{\mu_2}{2}\log\frac{|\bbk_X|}{|\bbd|}
\label{t_plus_lower_bound}
\end{align}
Finally, we note that
$(\bbd_1=\bbsigma_1,\bbd_2=\bbsigma_2\left(\bbk_X^{-1}+\bbsigma_2^{-1}-\bbd^{-1}\right)\bbsigma_2)\in\mathcal{D}^+(\bbd_1,\bbd_2)$
attains the lower bound for ${\rm T^+}$ in
(\ref{t_plus_lower_bound}); which completes the proof of
Corollary~\ref{Corollary_after_assumptions}.

\subsection{Proof of Corollary~\ref{Corollary_after_assumptions_chen}}
\label{sec:proof_of_corollary_after_assumptions_chen}

To obtain an outer bound for ${\rm T^-}$, we consider the
following $(\bbd_1,\bbd_2)$ pair
\begin{align}
\bbd_1&=\bbsigma_1\\
\bbd_2&=\frac{\mu_2}{\mu_1}\bbsigma_2\left(\bbk_X^{-1}+\bbsigma_2^{-1}\right)\bbsigma_2
\end{align}
which is feasible, i.e.,
$(\bbd_1,\bbd_2)\in\mathcal{D}^-(\bbd_1,\bbd_2)$. (To show that
$\bbd_2$ is feasible, we use (\ref{second_assumption}).)
Consequently, using this pair of matrices in the cost function of
${\rm T^-}$, we get the following upper bound for ${\rm T^-}$
\begin{align}
{\rm T^-}&\leq
\frac{\mu_2}{2}\log\frac{|\bbsigma_2^{-1}|}{\left|\frac{\mu_2}{\mu_1}\left(\bbk_X^{-1}+\bbsigma_2^{-1}\right)\right|}+\frac{\mu_2}{2}\log\frac{|\bbk_X|}{|\bbd|}
+\frac{\mu_1-\mu_2}{2}\log\frac{\left|\frac{\mu_1}{\mu_1-\mu_2}\left(\bbk_X^{-1}+\bbsigma_2^{-1}\right)^{-1}\right|}{|\bbd|}\\
&=\frac{\mu_2}{2}\log\frac{|\bbsigma_2^{-1}|}{|\bbk_X^{-1}+\bbsigma_2^{-1}-\bbd^{-1}|}
+\frac{\mu_2}{2}\log\frac{|\bbk_X^{-1}+\bbsigma_2^{-1}-\bbd^{-1}|}
{\left|\frac{\mu_2}{\mu_1}\left(\bbk_X^{-1}+\bbsigma_2^{-1}\right)\right|}+\frac{\mu_2}{2}\log\frac{|\bbk_X|}{|\bbd|}\nonumber
\\
&\quad
+\frac{\mu_1-\mu_2}{2}\log\frac{\left|\frac{\mu_1}{\mu_1-\mu_2}\left(\bbk_X^{-1}+\bbsigma_2^{-1}\right)^{-1}\right|}{|\bbd|}\\
&={\rm T^+}
+\frac{\mu_2}{2}\log\frac{|\bbk_X^{-1}+\bbsigma_2^{-1}-\bbd^{-1}|}
{\left|\frac{\mu_2}{\mu_1}\left(\bbk_X^{-1}+\bbsigma_2^{-1}\right)\right|}
+\frac{\mu_1-\mu_2}{2}\log\frac{\left|\frac{\mu_1}{\mu_1-\mu_2}\left(\bbk_X^{-1}+\bbsigma_2^{-1}\right)^{-1}\right|}{|\bbd|}
\end{align}
which is the desired end result in
Corollary~\ref{Corollary_after_assumptions_chen}; completing the
proof.

\section{Proof of Theorem~\ref{theorem_r_d_parallel}}
\label{proof_of_theorem_r_d_parallel}

We prove Theorem~\ref{theorem_r_d_parallel} in two steps. In the
first step, we specialize the outer bound
in~\cite{Wagner_Outer_Bound} to the parallel model defined by the
following joint distribution
\begin{align}
p(x^M,\{y^M_{\ell}\}_{\ell=1}^L)=\prod_{m=1}^M
p(x_m)\prod_{\ell=1}^Lp(y_{\ell m}|x_m) \label{parallel_models}
\end{align}
Next, we evaluate the outer bound we obtain in the first step, and
show that it can be attained by the inner bound provided in
Theorem~\ref{theorem_inner}.

\subsection{A General Outer Bound}
First, we restate the outer bound in~\cite{Wagner_Outer_Bound} for
the parallel model satisfying (\ref{parallel_models}) as follows.
\begin{Theo}{\bf (\!\!\!\cite[Theorem 1]{Wagner_Outer_Bound})}
\label{theorem_outer_wagner_p}
 We have $\mathcal{R}^{p}(\{D_m\}_{m=1}^M)\subseteq
\mathcal{R}^{p-o}(\{D_m\}_{m=1}^M)$, where $\break
\mathcal{R}^{p-o}(\{D_m\}_{m=1}^M)$ is given by the union of rate
tuples $(R_1,\ldots,R_L)$ satisfying
\begin{align}
\sum_{\ell\in\mathcal{A}}R_\ell& \geq
I(X^M;\{U_\ell\}_{\ell\in\mathcal{A}}|\{U_\ell\}_{\ell\in\mathcal{A}^c})+\sum_{\ell\in\mathcal{A}}I(U_\ell;Y_\ell^M|X^M,W)
\label{rate_constraints_parallel_initial}
\end{align}
for all $\mathcal{A}\subseteq\{1,\ldots,L\}$, where the union is
over all $\{U_\ell\}_{\ell=1}^L$ satisfying
\begin{align}
p(x^M,\{y^M_\ell\}_{\ell=1}^L,\{u_\ell\}_{\ell=1}^L,w)=p(w)\prod_{m=1}^M
p(x_m)\prod_{\ell=1}^L p(y_{\ell m}|x_m)p(u_\ell|w,y_\ell^M)
\label{joint_distribution_parallel_initial}
\end{align}
and
\begin{align}
\mmse(X_m|U_1,\ldots,U_L)\leq D_m,\quad m=1,\ldots,M
\label{distortion_constraints_parallel_initial}
\end{align}
\end{Theo}

Next, we define the following auxiliary random variables
\begin{align}
U_{\ell m}&=U_\ell X^{m-1},\qquad \qquad\qquad
\ell=1,\ldots,L,~~m=1,\ldots,M
\label{aux_1}\\
W_m&=WX^{m-1}\{Y_{\ell,m+1}^M\}_{\ell=1}^L,\quad m=1,\ldots,M
\label{aux_2}
\end{align}
Using these auxiliary random variables, we will find lower bounds
for the rate constraints in
(\ref{rate_constraints_parallel_initial}). We start with the
following term
\begin{align}
I(X^M;\{U_{\ell}\}_{\ell\in\mathcal{A}}|\{U_{\ell}\}_{\ell\in\mathcal{A}^c})&=\sum_{m=1}^M
I(X_m;\{U_{\ell}\}_{\ell\in\mathcal{A}}|\{U_{\ell}\}_{\ell\in\mathcal{A}^c},X^{m-1})\\
&=\sum_{m=1}^M I(X_m;\{U_{\ell m
}\}_{\ell\in\mathcal{A}}|\{U_{\ell
m}\}_{\ell\in\mathcal{A}^c})\label{rate_constraints_parallel_initial_1}
\end{align}
Next, we consider the following term
\begin{align}
\lefteqn{I(U_{\ell};Y_{\ell}^M|X^M,W)=h(U_{\ell}|X^M,W)-h(U_{\ell}|X^M,W,Y_{\ell}^M)}\\
&\geq h(U_{\ell}|X^M,W,\{Y_j^M\}_{j=1,j\neq
\ell}^L)-h(U_{\ell}|X^M,W,Y_{\ell}^M)
\label{conditioning_reduces_p_1} \\
&=  h(U_{\ell}|X^M,W,\{Y_j^M\}_{j=1,j\neq
\ell}^L)-h(U_{\ell}|X^M,W,\{Y_j^M\}_{j=1}^L)
\label{joint_distribution_parallel_initial_implies_1} \\
&=  I(U_{\ell};Y_\ell^M|X^M,W,\{Y_j^M\}_{j=1,j\neq \ell}^L)\\
&=\sum_{m=1}^M I(U_{\ell};Y_{\ell m}|X^M,W,\{Y_j^M\}_{j=1,j\neq
\ell}^L,Y_{\ell,m+1}^M) \\
&= \sum_{m=1}^M h(Y_{\ell m}|X^M,W,\{Y_j^M\}_{j=1,j\neq
\ell}^L,Y_{\ell,m+1}^M)-h(Y_{\ell m}|X^M,W,\{Y_j^M\}_{j=1,j\neq
\ell}^L,Y_{\ell,m+1}^M,U_{\ell})\\
&=\sum_{m=1}^M h(Y_{\ell
m}|X^m,W,\{Y_{j,m+1}^M\}_{j=1}^L)-h(Y_{\ell
m}|X^M,W,\{Y_j^M\}_{j=1,j\neq \ell}^L,Y_{\ell,m+1}^M,U_{\ell})
\label{joint_distribution_parallel_initial_implies_2}\\
&\geq \sum_{m=1}^M h(Y_{\ell
m}|X^m,W,\{Y_{j,m+1}^M\}_{j=1}^L)-h(Y_{\ell
m}|X^m,W,\{Y_{j,m+1}^M\}_{j=1}^L,U_{\ell})
\label{conditioning_reduces_p_2} \\
&= \sum_{m=1}^M I(U_{\ell};Y_{\ell
m}|X^m,W,\{Y_{j,m+1}^M\}_{j=1}^L)\\
&= \sum_{m=1}^M I(U_{\ell m};Y_{\ell m}|X_m,W_m)
\label{def_auxs_imply}
\end{align}
where (\ref{conditioning_reduces_p_1}) follows from the fact that
conditioning cannot increase entropy,
(\ref{joint_distribution_parallel_initial_implies_1}) and
(\ref{joint_distribution_parallel_initial_implies_2}) come from
the following Markov chains
\begin{align}
U_\ell &\rightarrow W,Y_\ell^M \rightarrow
X^M,\{Y_j^M\}_{j=1,j\neq
\ell}^L \\
Y_{\ell m} &\rightarrow X_m \rightarrow
W,X^{m-1},X_{m+1}^M,\{Y_j^M\}_{j=1,j\neq \ell}^L,Y_{\ell,m+1}^M
\end{align}
respectively, which are consequences of the joint distribution in
(\ref{joint_distribution_parallel_initial}), and
(\ref{conditioning_reduces_p_2}) is due to the fact that
conditioning cannot increase entropy.

Next, we consider the distortion constraints in
(\ref{distortion_constraints_parallel_initial}) as follows
\begin{align}
D_m&\geq \mmse(X_m|\{U_\ell\}_{\ell=1}^L)\\
&\geq \mmse(X_m|\{U_\ell\}_{\ell=1}^L,X^{m-1})\\
&=\mmse(X_m|\{U_{\ell m}\}_{\ell=1}^L)
\label{distortion_constraints_parallel_initial_1}
\end{align}
where we use the fact that conditioning reduces MMSE.

Hence, using (\ref{rate_constraints_parallel_initial_1}) and
(\ref{def_auxs_imply}), the rate constraints in
Theorem~\ref{theorem_outer_wagner_p} can be expressed as
\begin{align}
\sum_{\ell\in\mathcal{A}}R_\ell\geq \sum_{m=1}^M I(X_m;\{U_{\ell m
}\}_{\ell\in\mathcal{A}}|\{U_{\ell
m}\}_{\ell\in\mathcal{A}^c})+\sum_{m=1}^M
\sum_{\ell\in\mathcal{A}} I(U_{\ell m};Y_{\ell m}|X_m,W_m)
\label{rate_constraints_parallel}
\end{align}
and the distortion constraints in
Theorem~\ref{theorem_outer_wagner_p} are
\begin{align}
\mmse(X_m|\{U_{\ell m}\}_{\ell=1}^L)\leq D_m
\label{distortion_constraints_parallel}
\end{align}
We note that the random variable tuples
\begin{align}
\left\{\left(X_m,\{Y_{\ell m},U_{\ell m}\}_{\ell=1}^L,W_m \right)
\right\}_{m=1}^M
\end{align}
might be correlated over the index $m$. However, neither the
expressions in the rate bounds given by
(\ref{rate_constraints_parallel}) nor the distortion constraints
in (\ref{distortion_constraints_parallel}) depend on the entire
joint distribution of $\left\{\left(X_m,\{Y_{\ell m},U_{\ell
m}\}_{\ell=1}^L,W_m \right) \right\}_{m=1}^M$. Instead, both the
expressions in the rate bounds given by
(\ref{rate_constraints_parallel}) and the distortion constraints
in (\ref{distortion_constraints_parallel}) depend only on the
distribution of $\left(X_m,\{Y_{\ell m},U_{\ell
m}\}_{\ell=1}^L,W_m \right)$ for each $m$ involved. Hence, without
loss of generality, we can assume that
\begin{align}
\left(X_m,\{Y_{\ell m},U_{\ell m}\}_{\ell=1}^L,W_m \right)\quad
{\rm and }\quad \left\{\left(X_j,\{Y_{\ell j},U_{\ell
j}\}_{\ell=1}^L,W_j \right) \right\}_{j=1,j\neq m}^M
\end{align}
are independent for all $m=1,\ldots,M$. Next, we note that the
joint distribution of \break $\left(X_m,\{Y_{\ell m},U_{\ell
m}\}_{\ell=1}^L,W_m \right)$ can be factorized as follows
\begin{align}
p(x_m,\{y_{\ell m},u_{\ell m}\}_{\ell=1}^L,w_m
)=p(x_m)p(w_m)\prod_{\ell=1}^L p(y_{\ell m}|x_m) p(u_{\ell
m}|y_{\ell m}, w_m) \label{factorization_parallel}
\end{align}
whose proof is given in
Appendix~\ref{proof_of_factorization_parallel}. In view of
(\ref{rate_constraints_parallel})-(\ref{distortion_constraints_parallel})
and (\ref{factorization_parallel}), we obtain the following outer
bound for the parallel model.
\begin{Theo}
\label{theorem_outer_parallel}
 We have $\mathcal{R}^{p}(\{D_m\}_{m=1}^M)\subseteq
\mathcal{R}^{p-o}(\{D_m\}_{m=1}^M)$, where $ \mathcal{R}^{
p-o}(\{D_m\}_{m=1}^M)$ is given by the union of rate tuples
$(R_1,\ldots,R_L)$ satisfying
\begin{align}
\sum_{\ell\in\mathcal{A}}R_\ell& \geq \sum_{m=1}^M I(X_m;\{U_{\ell
m}\}_{\ell\in\mathcal{A}}|\{U_{\ell
m}\}_{\ell\in\mathcal{A}^c})+\sum_{m=1}^M
\sum_{\ell\in\mathcal{A}} I(U_{\ell m};Y_{\ell m}|X_m,W_m)
\label{rate_constraints_parallel_final}
\end{align}
for all $\mathcal{A}\subseteq\{1,\ldots,L\}$, where the union is
over all $\{U_{\ell m}\}_{\forall \ell, \forall m}$ satisfying
\begin{align}
p(x^M,\{y^M_\ell\}_{\ell=1}^L,\{u_\ell\}_{\ell=1}^L,w)=\prod_{m=1}^M
 p(x_m) p(w_m)\prod_{\ell=1}^L p(y_{\ell
m}|x_m)p(u_{\ell m}|w_m,y_{\ell m})
\label{joint_distribution_parallel_final}
\end{align}
and
\begin{align}
\mmse(X_m|\{U_{\ell m}\}_{\ell=1}^L)\leq D_m,\quad m=1,\ldots,M
\label{distortion_constraints_parallel_final}
\end{align}
\end{Theo}

\subsection{Evaluation of the Outer Bound}
\label{sec:evaluation_of_the_outer_bound}

Now, we evaluate the outer bound in
Theorem~\ref{theorem_outer_parallel} for the parallel Gaussian
model, and show that it is attainable by the inner bound given in
Theorem~\ref{theorem_inner}. To this end, we note that following
the analysis in Section~\ref{sec:proof_of_outer_bound}, one can
evaluate the outer bound in Theorem~\ref{theorem_outer_parallel}
yielding the following outer bound for the parallel Gaussian
model.
\begin{Theo}
\label{theorem_outer_parallel_Gaussian} An outer bound for the
rate-distortion region $\mathcal{R}^{p}(\{D_m\}_{m=1}^M)$ of the
parallel Gaussian model is given by
$\tilde{\mathcal{R}}^{p}(\{D_m\}_{m=1}^M)$ which corresponds to
the union of rate tuples $(R_1,\ldots,R_L)$ satisfying
\begin{align}
\sum_{\ell\in\mathcal{A}}R_\ell\geq
\sum_{m=1}^M\frac{1}{2}\log^+\frac{1}{D_m}
\left(\frac{1}{\sigma_m^2}+\sum_{\ell\in\mathcal{A}^c}\frac{\sigma_{\ell
m}^2-D_{\ell m}}{\sigma_{\ell m}^4}\right)^{-1}
+\sum_{m=1}^M\sum_{\ell\in\mathcal{A}}\frac{1}{2}\log\frac{\sigma_{\ell
m}^2}{D_{\ell m}}
\end{align}
for all $\mathcal{A}\subseteq\{1,\ldots,L\}$, where the union is
over all $\{D_{\ell m}\}_{\forall \ell,\forall m}$ satisfying the
following constraints
\begin{align}
\left(\frac{1}{\sigma_m^2}+\sum_{\ell=1}^L\frac{\sigma_{\ell
m}^2-D_{\ell m}}{\sigma_{\ell m}^4}\right)^{-1}&\leq D_m,\quad m=1,\ldots,M \label{crucial_constraints}\\
0\leq D_{\ell m}&\leq \sigma_{\ell m}^2,\quad
\ell=1,\ldots,L,~~m=1,\ldots,M \label{not_so_crucial_constraint}
\end{align}
\end{Theo}

Next, we show that there is no loss of generality to assume that
the constraints in (\ref{crucial_constraints}) are satisfied with
equality. To prove this, we consider an alternative description of
the outer bound in Theorem~\ref{theorem_outer_parallel_Gaussian}
by means of the tangent hyperplanes. In other words, we consider
the following optimization problem
\begin{align}
\min_{(R_1,\ldots,R_L)\in\tilde{\mathcal{R}}^{P}(\{D_m\}_{m=1}^M)}~~\sum_{\ell=1}^L~\mu_\ell
R_\ell \label{tangent_hyperplanes_P_G}
\end{align}
where we assume $\mu_1\geq \ldots \geq \mu_L\geq 0$. Using the
analysis in Appendix~\ref{proof_of_tangent_line_bounds}, we can
express the optimization problem in
(\ref{tangent_hyperplanes_P_G}) as follows
\begin{align}
\lefteqn{\min_{(R_1,\ldots,R_L)\in\tilde{\mathcal{R}}^{P}(\{D_m\}_{m=1}^M)}~~\sum_{\ell=1}^L~\mu_\ell
R_\ell  }\hspace{1.5cm}\nonumber\\
&=\min_{\{D_{\ell m}\}_{\forall \ell,\forall
m}}~~\sum_{\ell=1}^{L-1}~~\frac{\mu_\ell-\mu_{\ell+1}}{2}
~\sum_{m=1}^M
\log^{+}\frac{1}{D_m}\left(\frac{1}{\sigma_m^2}+\sum_{j=\ell+1}^L~\frac{\sigma_{j
m}^2-D_{j m}}{\sigma_{j m}^4}\right)^{-1} \nonumber\\
&\qquad \qquad \qquad
+\sum_{\ell=1}^L\frac{\mu_\ell}{2}\sum_{m=1}^M
\log\frac{\sigma_{\ell m}^2}{D_{\ell
m}}+\frac{\mu_L}{2}\sum_{m=1}^M \log\frac{\sigma_m^2}{D_m}
\label{tangent_hyperplanes_P_G_alternative}\\
&=\min_{\{D_{\ell m}\}_{\forall \ell,\forall m}}\sum_{m=1}^M~
f_{m}(\{D_{\ell m}\}_{\ell=1}^L) \label{separable_functions_pre} \\
&=\sum_{m=1}^M \min_{\{D_{\ell m}\}_{\ell=1}^L}~f_m(\{D_{\ell
m}\}_{\ell=1}^L) \label{separable_functions}
\end{align}
where we define the function $f_m(\{D_{\ell m}\}_{\ell=1}^L)$ as
follows
\begin{align}
f_m(\{D_{\ell
m}\}_{\ell=1}^L)&=\sum_{\ell=1}^{L-1}~~\frac{\mu_\ell-\mu_{\ell+1}}{2}
\log^{+}\frac{1}{D_m}\left(\frac{1}{\sigma_m^2}+\sum_{j=\ell+1}^L~\frac{\sigma_{j
m}^2-D_{j m}}{\sigma_{j m}^4}\right)^{-1}
+\sum_{\ell=1}^L\frac{\mu_\ell}{2} \log\frac{\sigma_{\ell
m}^2}{D_{\ell m}}\nonumber\\
&\quad +\frac{\mu_L}{2} \log\frac{\sigma_m^2}{D_m}
\end{align}
and the feasible set of the minimizations in
(\ref{tangent_hyperplanes_P_G_alternative})-(\ref{separable_functions})
are defined by the constraints in
(\ref{crucial_constraints})-(\ref{not_so_crucial_constraint}).
Equation (\ref{separable_functions}) follows from the fact that
$f_m(\{D_{\ell m}\}_{\ell=1}^L)$ depends only on $\{D_{\ell
m}\}_{\ell=1}^L$ but not on $\{D_{\ell j}\}_{\ell=1}^L,~j\neq m$.

Next, we note that each minimization
\begin{align}
\min_{\{D_{\ell m}\}_{\ell=1}^L}~f_m(\{D_{\ell m}\}_{\ell=1}^L)
\end{align}
is identical to the optimization problem we encounter for the
scalar Gaussian model in Section~\ref{sec:scalar_Gaussian_model},
and hence, the minimum is attained by those $\{D_{\ell
m}\}_{\ell=1}^L$ that satisfy the constraint in
(\ref{crucial_constraints}) with equality. This implies that the
outer bound in Theorem~\ref{theorem_outer_parallel_Gaussian} is
attainable; completing the proof of
Theorem~\ref{theorem_r_d_parallel}.

\subsection{Proof of (\ref{factorization_parallel})}
\label{proof_of_factorization_parallel}

We first note that
\begin{align}
p(x_m,\{y_{\ell m},u_{\ell
m}\}_{\ell=1}^L,w_m)&=p(x_m)p(w_m)\left(\prod_{\ell=1}^L p(y_{\ell
m}|x_m)\right) p(\{u_{\ell m}\}_{\ell=1}^L|x_m,w_m,\{y_{\ell
m}\}_{\ell=1}^L) \label{joint_distribution_p}
\end{align}
where we use the fact that $(X_m,\{Y_{\ell m}\}_{\ell=1}^L)$ and
$W_m=WX^{m-1}\{Y_{\ell,m+1}^M\}_{\ell=1}^L$ are independent, which
is a consequence of the joint distribution in
(\ref{joint_distribution_parallel_initial}). Next, we consider the
following term
\begin{align}
\lefteqn{p(\{u_{\ell m}\}_{\ell=1}^L|x_m,w_m,\{y_{\ell
m}\}_{\ell=1}^L)=\sum_{\forall
\{y_{\ell}^{m-1}\}_{\ell=1}^L}p(\{u_{\ell
m}\}_{\ell=1}^L,\{y_{\ell}^{m-1}\}_{\ell=1}^L|x_m,w_m,\{y_{\ell
m}\}_{\ell=1}^L)}\hspace{2cm}\\
& =\sum_{\forall \{y_{\ell}^{m-1}\}_{\ell=1}^L}
p(\{y_{\ell}^{m-1}\}_{\ell=1}^L|x_m,w_m,\{y_{\ell m}\}_{\ell=1}^L)
p(\{u_{\ell m}\}_{\ell=1}^L|x_m,w_m,\{y_{\ell}^{m}\}_{\ell=1}^L)
\label{cond_independence}
\end{align}
where the first term in the summation is
\begin{align}
p(\{y_{\ell}^{m-1}\}_{\ell=1}^L|x_m,w_m,\{y_{\ell
m}\}_{\ell=1}^L)&= \prod_{\ell=1}^L
p(y_{\ell}^{m-1}|x_m,w_m,\{y_{\ell
m}\}_{\ell=1}^L,\{y_{j}^{m-1}\}_{j=1}^{\ell-1})\label{first_MC}\\
&=\prod_{\ell=1}^L p(y_{\ell}^{m-1}|w_m) \label{second_MC}\\
&=\prod_{\ell=1}^{L}p(y_{\ell}^{m-1}|w_m,y_{\ell m})
\label{third_MC}
\end{align}
where (\ref{second_MC})-(\ref{third_MC}) come from the following
Markov chain
\begin{align}
Y_{\ell}^{m-1} \rightarrow W_m \rightarrow X_m,\{Y_{\ell
m}\}_{\ell=1}^L,\{Y_{j}^{m-1}\}_{j=1}^{\ell-1}
\end{align}
which is a consequence of the definition of $W_m$ and the joint
distribution in (\ref{joint_distribution_parallel_initial}).

Next, we consider the second term in the summation given by
(\ref{cond_independence}) as follows
\begin{align}
p(\{u_{\ell
m}\}_{\ell=1}^L|x_m,w_m,\{y_{j}^{m}\}_{j=1}^L)&=\prod_{\ell=1}^L
p(u_{\ell m}|x_m,w_m,\{y_{j}^{m}\}_{j=1}^L,\{u_{j
m}\}_{j=1}^{\ell-1})\\
&=\prod_{\ell=1}^L p(u_{\ell m}|w_m,y_{\ell}^{m})
\label{4th_MC_implies}
\end{align}
where (\ref{4th_MC_implies}) comes from the following Markov chain
\begin{align}
U_{\ell m}\rightarrow W_m,Y_{\ell}^{m}\rightarrow
X_m,\{Y_{j}^{m}\}_{j=1,j\neq \ell}^L,\{U_{j m}\}_{j=1}^{\ell-1}
\end{align}
which is also a consequence of the definition of $W_m$ and the
joint distribution in (\ref{joint_distribution_parallel_initial}).

Using (\ref{third_MC}) and (\ref{4th_MC_implies}) in
(\ref{cond_independence}), we get
\begin{align}
p(\{u_{\ell m}\}_{\ell=1}^L|x_m,w_m,\{y_{\ell m}\}_{\ell=1}^L) &
=\sum_{\forall \{y_{\ell}^{m-1}\}_{\ell=1}^L} \prod_{\ell=1}^L
p(y_{\ell}^{m-1}|w_m,y_{\ell m}) p(u_{\ell m}|w_m,y_{\ell}^{m})\\
&=\sum_{\forall \{y_{\ell}^{m-1}\}_{\ell=1}^L} \prod_{\ell=1}^L p(y_{\ell}^{m-1},u_{\ell m}|w_m,y_{\ell m}) \\
&=\prod_{\ell=1}^L p(u_{\ell m}|w_m,y_{\ell m})
\end{align}
using which in (\ref{joint_distribution_p}), we get
\begin{align}
p(x_m,\{y_{\ell m},u_{\ell
m}\}_{\ell=1}^L,w_m)&=p(x_m)p(w_m)\prod_{\ell=1}^L p(y_{\ell
m}|x_m) p(u_{\ell m}|w_m,y_{\ell m})
\end{align}
which is the desired result in (\ref{factorization_parallel});
completing the proof.

\section{Proof of Corollary~\ref{Corollary_after_assumptions_p}}
\label{proof_of_Corollary_after_assumptions_p}

From the analysis in
Appendix~\ref{sec:evaluation_of_the_outer_bound}, when $\mu_1\geq
\mu_2\geq 0$, we have
\begin{align}
\min_{(R_1,R_2)\in\mathcal{R}^{p}(D_{1},D_{2})}~\mu_1 R_1+\mu_2
R_2&=\sum_{m=1}^2 \min_{(D_{1m},D_{2m})\in\mathcal{D}_m}
f_m(D_{1m},D_{2m}) \label{tangent_hyperplanes_2}
\end{align}
where the function $f_m(D_{1m},D_{2m})$ is given by
\begin{align}
f_m (D_{1m},D_{2m})=\sum_{\ell=1}^2
\frac{\mu_\ell}{2}\log\frac{\sigma_{\ell m}^2}{D_{\ell
m}}+\frac{\mu_2}{2} \log\frac{\sigma_m^2}{D_m}
+\frac{\mu_1-\mu_2}{2} \log\frac{1}{D_m}
\left(\frac{1}{\sigma_m^2}+\frac{\sigma_{2m}^2-D_{2m}}{\sigma_{2m}^4}\right)^{-1}
\end{align}
and the set $\mathcal{D}_m$ consists of $(D_{1m},D_{2m})$ pairs
satisfying
\begin{align}
\frac{1}{\sigma_m^2}+\sum_{\ell=1}^2\frac{\sigma_{\ell
m}^2-D_{\ell m}}{\sigma_{\ell m}^4}&=\frac{1}{D_m} \\
0\leq D_{\ell m}&\leq \sigma_{\ell m}^2,\quad \ell=1,2
\end{align}
Next, we define the function $\tilde{f}_2(D_{12},D_{22})$ as
\begin{align}
\tilde{f}_2(D_{12},D_{22})&=\sum_{\ell=1}^2
\frac{\mu_\ell}{2}\log\frac{\sigma_{\ell 2}^2}{D_{\ell
2}}+\frac{\mu_2}{2}\log\frac{\sigma_2^2}{D_2}
\end{align}
and the set $\tilde{\mathcal{D}}_2$ as the union of
$(D_{12},D_{22})$ satisfying
\begin{align}
\left(\frac{1}{\sigma_2^2}+\sum_{\ell=1}^2 \frac{\sigma_{\ell
2}^2-
D_{\ell 2}}{\sigma_{\ell 2}^4}\right)^{-1}&\leq D_2\\
0\leq
D_{12}&\leq \sigma_{12}^2
\end{align}
We note the following facts
\begin{align}
f_{2}(D_{12},D_{22})&\geq \tilde{f}_2 (D_{12},D_{22}),\quad
\forall (D_{12},D_{22})\in\mathcal{D}_2 \label{fact_1}\\
\mathcal{D}_2&\subseteq \tilde{\mathcal{D}}_2 \label{fact_2}
\end{align}
Next, we consider the following optimization problem
\begin{align}
\min_{(D_{12},D_{22})\in\mathcal{D}_2}~f_2(D_{12},D_{22})&\geq
\min_{(D_{12},D_{22})\in\mathcal{D}_2}~\tilde{f}_2(D_{12},D_{22})\label{fact_1_implies}
\\
&\geq
\min_{(D_{12},D_{22})\in\tilde{\mathcal{D}}_2}~\tilde{f}_2(D_{12},D_{22})
\label{fact_2_implies}
\end{align}
where (\ref{fact_1_implies})-(\ref{fact_2_implies}) follow from
(\ref{fact_1})-(\ref{fact_2}), respectively. We note that the
optimization problem in (\ref{fact_2_implies}) is the scalar
version of the optimization problem we consider in
Appendix~\ref{sec:proof_of_corollary_after_assumptions}. Using the
result from
Appendix~\ref{sec:proof_of_corollary_after_assumptions}, we have
\begin{align}
\min_{(D_{12},D_{22})\in\tilde{\mathcal{D}}_2}~\tilde{f}_2(D_{12},D_{22})=
\frac{\mu_2}{2}\log\frac{\sigma_2^2}{D_2}
+\frac{\mu_2}{2}\log\frac{1}{\sigma_{22}^2}\left(\frac{1}{\sigma_2^2}+\frac{1}{\sigma_{22}^2}-\frac{1}{D_2}\right)^{-1}
\label{dummy_optimization}
\end{align}
Next, we note that by setting
$(D^*_{12}=\sigma_{12}^2,D^*_{22}=\sigma_{22}^4(1/\sigma_2^2+1/\sigma_{22}^2-1/D_2))\in\mathcal{D}_{2}$,
we get
\begin{align}
f_2(D_{12}^*,D_{22}^*)=\min_{(D_{12},D_{22})\in\tilde{\mathcal{D}}_2}~\tilde{f}_2(D_{12},D_{22})
\end{align}
using which, and (\ref{dummy_optimization}) in
(\ref{tangent_hyperplanes_2}), we get
\begin{align}
\min_{(R_1,R_2)\in\mathcal{R}^p(D_{1},D_{2})}~\mu_1 R_1+\mu_2
R_2&=\min_{(D_{11},D_{21})\in\mathcal{D}_1}
f_1(D_{11},D_{21})\nonumber\\
&\qquad + \frac{\mu_2}{2}\log\frac{\sigma_2^2}{D_2}
+\frac{\mu_2}{2}\log\frac{1}{\sigma_{22}^2}\left(\frac{1}{\sigma_2^2}+\frac{1}{\sigma_{22}^2}-\frac{1}{D_2}\right)^{-1}
\end{align}
which is the desired result in
Corollary~\ref{Corollary_after_assumptions_p}; completing the
proof.

\section{Proof of Corollary~\ref{corollary_after_assumptions_ours_p}}
\label{proof_of_corollary_after_assumptions_ours_p}

Using Corollary~\ref{corollary_our_outer_bound}, our outer bound
for the parallel Gaussian model can be expressed as follows.
\begin{align}
{\rm T^+}&=\min_{(R_1,R_2)\in\mathcal{R}^o(D_1,D_2)}~\mu_1
R_1+\mu_2 R_2\\
&=\min_{(\bbd_1,\bbd_2,\bbd)}~\sum_{\ell=1}^2\frac{\mu_\ell}{2}\log\frac{|\bbsigma_\ell|}{|\bbd_\ell|}+\frac{\mu_2}{2}\log\frac{|\bbk_X|}{|\bbd|}
\nonumber\\
&\qquad \qquad \qquad
+\frac{\mu_1-\mu_2}{2}\log^{+}\frac{\left|\left(\bbk_X^{-1}+\bbsigma_2^{-1}-\bbsigma_2^{-1}\bbd_2\bbsigma_2^{-1}\right)^{-1}\right|}{|\bbd|}
\end{align}
where $(\bbd_1,\bbd_2,\bbd)$ are subject to the following
constraints
\begin{align}
\left(\bbk_X^{-1}+\sum_{\ell=1}^2\bbsigma_\ell^{-1}-\sum_{\ell=1}^2\bbsigma_{\ell}^{-1}\bbd_\ell\bbsigma_\ell^{-1}\right)^{-1}&\preceq
\bbd \\
\bzero\preceq \bbd_\ell &\preceq \bbsigma_\ell,\quad \ell=1,2 \\
\bbd_{mm}&\leq D_m,\quad m=1,2
\end{align}
where $\bbd_{mm}$ denotes the $m$th diagonal element of $\bbd$. By
restricting $(\bbd_1,\bbd_2,\bbd)$ to be diagonal, we have
\begin{align}
{\rm T^+}& \leq \min_{\{D_{\ell m}\}_{\forall \ell, \forall
m}}~\sum_{m=1}^2\frac{\mu_1}{2}\log\frac{\sigma_{1m}^2}{D_{1m}}+\frac{\mu_2}{2}\log\frac{\sigma_{2m}^2}{D_{2m}}+\frac{\mu_2}{2}\log\frac{\sigma_m^2}{D_m}
\nonumber\\
&\qquad \qquad \qquad  +\frac{\mu_1-\mu_2}{2}\log^{+}\prod_{m=1}^2
\frac{1}{D_m}\left(\frac{1}{\sigma_m^2}+\frac{1}{\sigma_{2m}^2}-\frac{D_{2m}}{\sigma_{2m}^4}\right)^{-1}
\label{tangent_to_ours}
\end{align}
where $\{D_{\ell m}\}_{\forall \ell, \forall m}$ are subject to
the following constraints
\begin{align}
\left(\frac{1}{\sigma_{m}^2}+\sum_{\ell=1}^2\frac{1}{\sigma_{\ell
m}^2}-\frac{D_{\ell m}}{\sigma_{\ell m}^4}\right)^{-1}&\leq
D_m,\quad m=1,2 \label{constraint_1}\\
0\leq D_{\ell m } &\leq \sigma_{\ell m}^2,\quad \ell=1,2~~m=1,2
\label{constraint_2}
\end{align}
Next, we set
\begin{align}
D_{12}&=\sigma_{12}^2 \label{select_1}\\
D_{22}&=\frac{\mu_2}{\mu_1}\sigma_{22}^4
\left(\frac{1}{\sigma_2^2}+\frac{1}{\sigma_{22}^2}\right)
\label{select_2}
\end{align}
which are feasible, i.e., satisfy the constraints in
(\ref{constraint_1})-(\ref{constraint_2}), due to the assumptions
in (\ref{second_assumption_p})-(\ref{third_assumption_p}). Next,
we note the following
\begin{align}
\prod_{m=1}^2
\frac{1}{D_m}\left(\frac{1}{\sigma_m^2}+\frac{1}{\sigma_{2m}^2}-\frac{D_{2m}}{\sigma_{2m}^4}\right)^{-1}
&=\frac{1}{D_1}\left(\frac{1}{\sigma_1^2}+\frac{1}{\sigma_{21}^2}-\frac{D_{21}}{\sigma_{21}^4}\right)^{-1}\frac{1}{D_2}\frac{\mu_1}{\mu_1-\mu_2}
\left(\frac{1}{\sigma_2^2}+\frac{1}{\sigma_{22}^2}\right)^{-1}\label{selections_imply}\\
&\geq
\frac{1}{D_1}\left(\frac{1}{\sigma_1^2}+\frac{1}{\sigma_{21}^2}\right)^{-1}\frac{1}{D_2}\frac{\mu_1}{\mu_1-\mu_2}
\left(\frac{1}{\sigma_2^2}+\frac{1}{\sigma_{22}^2}\right)^{-1}\label{non_zero}
\\
&> 1 \label{fourth_assumption_p_implies}
\end{align}
where, in (\ref{selections_imply}), we use
(\ref{select_1})-(\ref{select_2}), (\ref{non_zero}) follows from
the fact that $D_{21}\geq 0$, and
(\ref{fourth_assumption_p_implies}) is due to the assumption in
(\ref{fourth_assumption_p}). Hence, using
(\ref{select_1})-(\ref{select_2}) and (\ref{non_zero}) in
(\ref{tangent_to_ours}), we get
\begin{align}
{\rm T^+}& \leq \min_{(D_{11},D_{21}) \in\hat{\mathcal{D}}_1}~
f_1(D_{11},D_{21}) +\frac{\mu_2}{2}\log\frac{\sigma_2^2}{D_2}
+\frac{\mu_2}{2}\log\frac{\mu_1}{\mu_2}\frac{1}{\sigma_{22}^2}\left(\frac{1}{\sigma_2^2}+\frac{1}{\sigma_{22}^2}\right)^{-1}\nonumber\\
&\qquad \qquad \qquad
+\frac{\mu_1-\mu_2}{2}\log\frac{\mu_1}{\mu_1-\mu_2}\frac{1}{D_2}\left(\frac{1}{\sigma_2^2}+\frac{1}{\sigma_{22}^2}\right)^{-1}
\label{further_upper_bound}
\end{align}
where the set $\hat{\mathcal{D}}_1$ is defined as the union of
$(D_{11},D_{21})$ pairs satisfying
\begin{align}
\left(\frac{1}{\sigma_1^2}+\sum_{\ell=1}^2 \frac{1}{\sigma_{\ell 1}^2}-\frac{D_{\ell 1}}{\sigma_{\ell 1}^4}\right)^{-1}&\leq D_1\\
0\leq D_{\ell 1}&\leq \sigma_{\ell 1}^2
\end{align}
We note that $\mathcal{D}_1\subseteq \hat{\mathcal{D}}_1$, where
$\mathcal{D}_1$ is the region defined in
Corollary~\ref{corollary_after_assumptions_ours_p}. Hence, using
this fact in (\ref{further_upper_bound}), we get
\begin{align}
{\rm T^+}&\leq \min_{(D_{11},D_{21}) \in\mathcal{D}_1}~
f_1(D_{11},D_{21}) +\frac{\mu_2}{2}\log\frac{\sigma_2^2}{D_2}
+\frac{\mu_2}{2}\log\frac{\mu_1}{\mu_2}\frac{1}{\sigma_{22}^2}\left(\frac{1}{\sigma_2^2}+\frac{1}{\sigma_{22}^2}\right)^{-1}\nonumber\\
&\qquad \qquad \qquad
+\frac{\mu_1-\mu_2}{2}\log\frac{\mu_1}{\mu_1-\mu_2}\frac{1}{D_2}\left(\frac{1}{\sigma_2^2}+\frac{1}{\sigma_{22}^2}\right)^{-1}
\end{align}
which is the desired result in
Corollary~\ref{corollary_after_assumptions_ours_p}; completing the
proof.

\section{Proof of Lemma~\ref{lemma_estimation}}
\label{proof_of_lemma_estimation}

We first note the following Markov chain
\begin{align}
(U_j,\bby_j)\rightarrow (W,\bbx) \rightarrow (U_{[1:L]\backslash
j},\bby_{[1:L]\backslash j}) \label{important_MC}
\end{align}
whose proof is given in Appendix~\ref{proof_of_MC}. Next, we note
that
\begin{align}
E\left[\bbs_{\mathcal{A}^c}|\bbx,\{U_\ell\}_{\ell\in\mathcal{A}^c},W\right]&=
\sum_{\ell\in\mathcal{A}^c}\bba_\ell
E\left[\bby_\ell|\bbx,\{U_\ell\}_{\ell\in\mathcal{A}^c},W\right]\\
&=\sum_{\ell\in\mathcal{A}^c}\bba_\ell
E\left[\bby_\ell|\bbx,U_\ell,W\right] \label{estimators_imply}
\end{align}
where (\ref{estimators_imply}) follows from the Markov chain in
(\ref{important_MC}). Now, we consider
$\mmse(\bbs_{\mathcal{A}^c}|\bbx,\{U_\ell\}_{\ell\in\mathcal{A}^c},W)$
as follows
\begin{align}
\lefteqn{\mmse(\bbs_{\mathcal{A}^c}|\bbx,\{U_\ell\}_{\ell\in\mathcal{A}^c},W)}\nonumber\\
&=
E\left[\Big(\bbs_{\mathcal{A}^c}-E\big[\bbs_{\mathcal{A}^c}|\bbx,\{U_\ell\}_{\ell\in\mathcal{A}^c},W\big]\Big)
\Big(\bbs_{\mathcal{A}^c}-E\big[\bbs_{\mathcal{A}^c}|\bbx,\{U_\ell\}_{\ell\in\mathcal{A}^c},W\big]\Big)^\top\right]\\
&= E\left[\Big(\sum_{\ell\in\mathcal{A}^c} \bba_\ell
\big(\bby_\ell-E\left[\bby_\ell|\bbx,U_\ell,W\right]\big)\Big)
\Big(\sum_{\ell\in\mathcal{A}^c} \bba_\ell
\big(\bby_\ell-E\left[\bby_\ell|\bbx,U_\ell,W\right]\big)\Big)^\top\right]
\label{estimators_imply_implies}
\\
&=\sum_{\ell\in\mathcal{A}^c}\bba_\ell
\mmse(\bby_\ell|\bbx,U_\ell,W)\bba_\ell^\top \nonumber\\
&\quad +
\sum_{i\in\mathcal{A}^c}\sum_{\substack{j\in\mathcal{A}^c\\j\neq
i}} \bba_i E\left[\big(\bby_i-E\left[\bby_i|\bbx,U_i,W\right]\big)
\big(\bby_j-E\left[\bby_j|\bbx,U_j,W\right]\big)^\top\right]\bba_j^\top
\label{cross-terms}
\end{align}
where (\ref{estimators_imply_implies}) is due to
(\ref{estimators_imply}). Next, we consider the cross-terms in
(\ref{cross-terms}) as follows
\begin{align}
\lefteqn{E\left[\big(\bby_i-E\left[\bby_i|\bbx,U_i,W\right]\big)
\big(\bby_j-E\left[\bby_j|\bbx,U_j,W\right]\big)^\top\right]}\nonumber\\
&= E\left[E\left[\big(\bby_i-E\left[\bby_i|\bbx,U_i,W\right]\big)
\big(\bby_j-E\left[\bby_j|\bbx,U_j,W\right]\big)^\top|\bbx,W\right]\right]\\
&=
E\left[E\left[\big(\bby_i-E\left[\bby_i|\bbx,U_i,W\right]\big)|\bbx,W\right]
E\left[\big(\bby_j-E\left[\bby_j|\bbx,U_j,W\right]\big)^\top|\bbx,W\right]\right]
\label{important_MC_implies_again} \\
&=\bzero \label{zero_cross_terms}
\end{align}
where (\ref{important_MC_implies_again}) is due to the Markov
chain in (\ref{important_MC}). Using (\ref{zero_cross_terms}) in
(\ref{cross-terms}), we get
\begin{align}
\mmse(\bbs_{\mathcal{A}^c}|\bbx,\{U_\ell\}_{\ell\in\mathcal{A}^c},W)
&=\sum_{\ell\in\mathcal{A}^c}\bba_\ell
\mmse(\bby_\ell|\bbx,U_\ell,W)\bba_\ell^\top
\end{align}
which completes the proof of Lemma~\ref{lemma_estimation}.

\subsection{Proof of (\ref{important_MC})} \label{proof_of_MC}

We first consider the joint distribution in
(\ref{joint_distribution}) as follows
\begin{align}
p(\bx,\{\by_\ell,u_\ell\}_{\ell=1}^L,w)=p(\bx)p(w)\left(\prod_{\substack{\ell=1\\
\ell \neq j}}^L p(\by_\ell|\bx)p(u_\ell|\by_\ell,w)\right)
p(\by_j|\bx) p(u_j|\by_j,w) \label{factorization}
\end{align}
which implies
\begin{align}
U_\ell\rightarrow (\bby_\ell,W)\rightarrow \bbx,\quad
\ell=1,\ldots,L\label{dummy_MC_x}
\end{align}
Next, we note that
\begin{align}
p(\by_j|\bx) p(u_j|\by_j,w)&=p(\by_j|\bx) p(u_j|\by_j,w,\bx)
\label{dummy_MC_x_implies} \\
&=p(\by_j|\bx) \frac{p(u_j,\by_j,w,\bx)}{p(\by_j,\bx)p(w)}
\label{independence_implies} \\
&= \frac{p(u_j,\by_j,w,\bx)}{p(\bx)p(w)}\\
&=
\frac{p(u_j,\by_j,w,\bx)}{p(w,\bx)}\label{independence_implies_1}\\
&=p(u_j,\by_j|w,\bx) \label{intermediate_result}
\end{align}
where (\ref{dummy_MC_x_implies}) comes from the Markov chain in
(\ref{dummy_MC_x}), (\ref{independence_implies}) and
(\ref{independence_implies_1}) follow from the fact that
$(\bbx,\bby_j)$ and $W$ are independent which is a consequence of
the factorization in (\ref{factorization}). Using
(\ref{intermediate_result}) in (\ref{factorization}), we get
\begin{align}
p(\bx,\{\by_\ell,u_\ell\}_{\ell=1}^L,w)=p(\bx)p(w)\left(\prod_{\substack{\ell=1\\
\ell \neq j}}^L p(\by_\ell|\bx)p(u_\ell|\by_\ell,w)\right)
p(u_j,\by_j|\bx,w)
\end{align}
which implies the Markov chain in (\ref{important_MC}); completing
the proof.

\section{Proof of Theorem~\ref{theorem_outer_general}}
\label{proof_of_theorem_outer_general}

The singular value decomposition of the matrices
$\{\bbh_\ell\}_{\ell=1}^L$ are given by
\begin{align}
\bbh_\ell&=\bbu_\ell \bblambda_\ell \bbv_\ell^\top,\quad
\ell=1,\ldots,L
\end{align}
where $\{\bbu_\ell\}_{\ell=1}^L$ and $\{\bbv_\ell\}_{\ell=1}^L$
are orthonormal matrices. Next, we show that without loss of
generality, we can assume that $\{\bbh_\ell\}_{\ell=1}^L$ are
square matrices. To this end, we define the following observations
\begin{align}
\bar{\bby}_\ell&=\bbu_\ell^\top\bby_\ell\\
&=\bblambda_\ell\bbv_\ell \bbx+\bar{\bbn}_\ell
\label{equivalent_model}
\end{align}
where $\bar{\bbn}_\ell$ is again a zero-mean Gaussian random
vector with an identity covariance matrix. We note that the
rate-distortion region for the observations
$\{\bar{\bby}_\ell\}_{\ell=1}^L$ is identical to the
rate-distortion region for the observations
$\{\bby_\ell\}_{\ell=1}^L$, since we obtain the observations
$\{\bar{\bby}_\ell\}_{\ell=1}^L$ from $\{\bby_\ell\}_{\ell=1}^L$
by an invertible transformation. Now, we show that there is no
loss of generality to assume that the matrices
$\{\bbh_\ell\}_{\ell=1}^L$ are square matrices. Assume that
$\bbh_\ell$ is an $r_\ell\times M$ matrix. Hence, $\bblambda_\ell$
is also an $r_\ell\times M$ matrix. First, consider $r_\ell >M$.
In this case, $r_\ell-M$ entries of $\bar{\bby}_{\ell}$ consists
of only noise. Since the noise $\bar{\bbn}_\ell$ is i.i.d., we can
drop these $r_\ell-M$ entries of the observation $\bar{\bby}_\ell$
without altering the rate-distortion region. Hence, when
$r_\ell>M$, there is an equivalent model with the same
rate-distortion region and $r_\ell=M$. Next, assume $r_\ell <M$.
In this case, we can add $M-r_\ell$ i.i.d. noise entries to the
observation $\bar{\bby}_\ell$ without altering the rate-distortion
region. Hence, when $r_\ell<M$, there is also an equivalent model
with the same rate-distortion region and $r_\ell=M$. Consequently,
from now on, we assume that $r_1=\ldots=r_L=M$.

Next, we define
\begin{align}
\bbh_{\ell,\alpha}&=\bbu_\ell (\bblambda_\ell+\alpha \bbi)
\bbv_\ell^\top,\quad \ell=1,\ldots,L
\end{align}
where $\alpha>0$. We note that $\{\bbh_{\ell,\alpha}\}_{\ell=1}^L$
are invertible, i.e., $\{\bbh_{\ell,\alpha}^{-1}\}_{\ell=1}^L$
exist, and in particular, we have
\begin{align}
\bbh_{\ell,\alpha}^{-1}=\bbv_\ell (\bblambda_\ell+\alpha
\bbi)^{-1}\bbu_\ell^\top,\quad \ell=1,\ldots,L
\end{align}
Using $\{\bbh_{\ell,\alpha}\}_{\ell=1}^L$, we define an {\it
enhanced model} as follows
\begin{align}
\bby_{\ell,\alpha}=\bbh_{\ell,\alpha}\bbx+\bbn_\ell,\quad
\ell=1,\ldots,L \label{enhanced_model}
\end{align}
Using these enhanced observations in (\ref{enhanced_model}), we
can rewrite the original observations in (\ref{general_form}) as
follows
\begin{align}
\bby_\ell=\bbh_\ell\bbh_{\ell,\alpha}^{-1}\bby_{\ell,\alpha}+\tilde{\bbn}_{\ell},\quad
\ell=1,\ldots,L \label{decomposition}
\end{align}
where $\tilde{\bbn}_\ell$ is a zero-mean Gaussian random vector,
and independent of $\{\bby_{\ell,\alpha}\}_{\ell=1}^L$ and
$\{\tilde{\bbn}_{j}\}_{j=1,j\neq \ell}^L$. The decomposition in
(\ref{decomposition}) is possible, since we have
\begin{align}
(\bbh_{\ell}\bbh_{\ell,\alpha}^{-1})(\bbh_{\ell}\bbh_{\ell,\alpha}^{-1})^\top&=\bbu_\ell
\bblambda_\ell^2 (\bblambda_\ell+\alpha \bbi)^{-2}
\bbu_\ell^\top\\
&\preceq \bbi
\end{align}
Moreover, due to the decomposition in (\ref{decomposition}), we
can assume that the following holds
\begin{align}
p(\bx,\{\by_{\ell,\alpha},\by_\ell\}_{\ell=1}^L)=p(\bx)\prod_{\ell=1}^L
p(\by_{\ell,\alpha}|\bx)p(\by_\ell|\by_{\ell,\alpha})
\end{align}
which implies that the original observations
$\{\bby_\ell\}_{\ell=1}^L$ are degraded versions the enhanced
observations $\{\bby_{\ell,\alpha}\}_{\ell=1}^L$. Consequently, we
have
\begin{align}
\mathcal{R}(\bbd)\subseteq \mathcal{R}_\alpha (\bbd)
\label{inclusion}
\end{align}
where $\mathcal{R}_{\alpha}(\bbd)$ denotes the rate-distortion
region for the enhanced model defined by (\ref{enhanced_model}).
Next, we note that the enhanced model defined by
(\ref{enhanced_model}) is equivalent to the following one
\begin{align}
\bar{\bby}_{\ell,\alpha}&=\bbh_{\ell,\alpha}^{-1}\bby_{\ell,\alpha}\\
&=\bbx+\bar{\bbn}_{\ell,\alpha},\quad \ell=1,\ldots,L
\label{enhanced_model_equi}
\end{align}
where the covariance matrix of $\bar{\bbn}_{\ell,\alpha}$ is given
by
\begin{align}
\bbsigma_{\ell,\alpha}=\left(\bbh_{\ell,\alpha}^\top\bbh_{\ell,\alpha}\right)^{-1},\quad
\ell=1,\ldots,L \label{noise_alpha}
\end{align}
Using Theorem~\ref{theorem_outer}, we can obtain an outer bound
for the rate-distortion region of the model defined by
(\ref{enhanced_model_equi}), which is equivalent to the enhanced
model given by (\ref{enhanced_model}). In particular, we have
$\mathcal{R}_\alpha (\bbd)\subseteq \mathcal{R}_\alpha^o(\bbd)$,
where $\mathcal{R}_\alpha^o(\bbd)$ is given by the union of rate
tuples $(R_1,\ldots,R_L)$ satisfying
\begin{align}
\sum_{\ell\in\mathcal{A}}R_\ell &\geq \frac{1}{2}\log^{+}
\frac{\left|\left(\bbk_X^{-1}+\sum_{\ell\in\mathcal{A}^c}\bbsigma_{\ell,\alpha}^{-1}-\sum_{\ell\in\mathcal{A}^c}\bbsigma_{\ell,\alpha}^{-1}
\tilde{\bbd}_\ell\bbsigma_{\ell,\alpha}^{-1}
\right)^{-1}\right|}{|\bbd|}+\sum_{\ell\in\mathcal{A}}\frac{1}{2}\log\frac{|\bbsigma_{\ell,\alpha}|}{|\tilde{\bbd}_\ell|}
\label{outer_bound_enhanced_1}
\end{align}
for all $\mathcal{A}\subseteq \{1,\ldots,L\}$, where the union is
over all $\{\tilde{\bbd}_\ell\}_{\ell=1}^L$ satisfying
\begin{align}
\left(\bbk_X^{-1}+\sum_{\ell=1}^L
\bbsigma_{\ell,\alpha}^{-1}-\sum_{\ell=1}^L\bbsigma_{\ell,\alpha}^{-1}
\tilde{\bbd}_\ell\bbsigma_{\ell,\alpha}^{-1} \right)^{-1}&\preceq
\bbd \label{outer_bound_enhanced_2}\\
\bzero \preceq \tilde{\bbd}_\ell &\preceq
\bbsigma_{\ell,\alpha},\quad \ell=1,\ldots,L
\label{outer_bound_enhanced_3}
\end{align}
Next, we set
$\bbd_\ell=\bbh_{\ell,\alpha}\tilde{\bbd}_{\ell}\bbh_{\ell,\alpha}^\top,~\ell=1,\ldots,L$,
using which in
(\ref{outer_bound_enhanced_1})-(\ref{outer_bound_enhanced_3}), we
can express the outer bound $\mathcal{R}_\alpha^o(\bbd)$ as the
union of rate tuples $(R_1,\ldots,R_L)$ satisfying
\begin{align}
\sum_{\ell\in\mathcal{A}}R_\ell &\geq \frac{1}{2}\log^{+}
\frac{\left|\left(\bbk_X^{-1}+\sum_{\ell\in\mathcal{A}^c}\bbh_{\ell,\alpha}^\top
(\bbi-\bbd_\ell)\bbh_{\ell,\alpha}
\right)^{-1}\right|}{|\bbd|}+\sum_{\ell\in\mathcal{A}}\frac{1}{2}\log\frac{1}{|\bbd_\ell|}
\label{outer_bound_enhanced_1_again}
\end{align}
for all $\mathcal{A}\subseteq \{1,\ldots,L\}$, where the union is
over all $\{\bbd_\ell\}_{\ell=1}^L$ satisfying
\begin{align}
\left(\bbk_X^{-1}+\sum_{\ell=1}^L\bbh_{\ell,\alpha}^\top
(\bbi-\bbd_\ell)\bbh_{\ell,\alpha}\right)^{-1}&\preceq
\bbd \label{outer_bound_enhanced_2_again}\\
\bzero \preceq \bbd_\ell &\preceq \bbi,\quad \ell=1,\ldots,L
\label{outer_bound_enhanced_3_again}
\end{align}
In view of (\ref{inclusion}), we have the following
\begin{align}
\mathcal{R}(\bbd) \subseteq \mathcal{R}_\alpha^o(\bbd),\quad
\forall \alpha >0
\end{align}
which implies that
\begin{align}
\mathcal{R}(\bbd) \subseteq \lim_{\alpha \rightarrow 0
}\mathcal{R}_\alpha^o(\bbd)
\end{align}
Hence, to obtain an outer bound for the rate-distortion region of
the general model defined by (\ref{general_form}), it is
sufficient to obtain the limiting region $\lim_{\alpha \rightarrow
0}\mathcal{R}_\alpha^o(\bbd)$. To this end, we introduce the
following lemma.
\begin{Lem}
\label{lemma_limiting_inverse} For all $\mathcal{A}^c\subseteq
\{1,\ldots,L\}$, we have
\begin{align}
\lim_{\alpha\rightarrow 0}
\left(\bbk_X^{-1}+\sum_{\ell\in\mathcal{A}^c}\bbh_{\ell,\alpha}^\top(\bbi-\bbd_\ell)\bbh_{\ell,\alpha}\right)^{-1}
&=\left(\bbk_X^{-1}+\sum_{\ell\in\mathcal{A}^c}\bbh_\ell^\top
(\bbi-\bbd_\ell)\bbh_\ell\right)^{-1}
\end{align}
\end{Lem}
The proof of Lemma~\ref{lemma_limiting_inverse} is given in
Appendix~\ref{proof_of_lemma_limiting_inverse}. Using this lemma
in
(\ref{outer_bound_enhanced_1_again})-(\ref{outer_bound_enhanced_3_again}),
we obtain the region $\lim_{\alpha\rightarrow 0}\mathcal{R}_\alpha
(\bbd)$ as the union of rate tuples $(R_1,\ldots,R_L)$ satisfying
\begin{align}
\sum_{\ell\in\mathcal{A}}R_\ell &\geq \frac{1}{2}\log^{+}
\frac{\left|\left(\bbk_X^{-1}+\sum_{\ell\in\mathcal{A}^c}\bbh_{\ell}^\top
(\bbi-\bbd_\ell)\bbh_{\ell}
\right)^{-1}\right|}{|\bbd|}+\sum_{\ell\in\mathcal{A}}\frac{1}{2}\log\frac{1}{|\bbd_\ell|}
\label{outer_bound_enhanced_1_again_f}
\end{align}
for all $\mathcal{A}\subseteq \{1,\ldots,L\}$, where the union is
over all $\{\bbd_\ell\}_{\ell=1}^L$ satisfying
\begin{align}
\left(\bbk_X^{-1}+\sum_{\ell=1}^L\bbh_{\ell}^\top
(\bbi-\bbd_\ell)\bbh_{\ell}\right)^{-1}&\preceq
\bbd \label{outer_bound_enhanced_2_again_f}\\
\bzero \preceq \bbd_\ell &\preceq \bbi,\quad \ell=1,\ldots,L
\label{outer_bound_enhanced_3_again_f}
\end{align}
which is the desired result in
Theorem~\ref{theorem_outer_general}; completing the proof.

\subsection{Proof of Lemma~\ref{lemma_limiting_inverse}}
\label{proof_of_lemma_limiting_inverse}

In the proof of Lemma~\ref{lemma_limiting_inverse}, we use the
following fact.
\begin{Lem}{\bf (\!\!\cite[page 258]{horn_johnson_book1})}
\label{lemma_matrix_inverse} Let $\bbc$ be a matrix satisfying
$\lim_{n\rightarrow \infty}\bbc^n=\bzero$. Then, we have
\begin{align}
(\bbi+\bbc)^{-1}=\sum_{n=0}^\infty (-1)^n\bbc^n
\end{align}
where $\bbc^0=\bbi$.
\end{Lem}
Next, we note that
\begin{align}
\sum_{\ell\in\mathcal{A}^c}\bbh_{\ell,\alpha}^\top
(\bbi-\bbd_\ell)\bbh_{\ell,\alpha}=\sum_{\ell\in\mathcal{A}^c}\bbh_\ell^\top
(\bbi-\bbd_\ell)\bbh_\ell+\bbm(\alpha) \label{def_M_alpha}
\end{align}
where $\lim_{\alpha\rightarrow 0}\bbm(\alpha)=\bzero$. We define
\begin{align}
\bbm_{\mathcal{A}^c}=\bbk_X^{-1}+\sum_{\ell\in\mathcal{A}^c}\bbh_\ell^\top
(\bbi-\bbd_\ell)\bbh_\ell \label{def_M_sum}
\end{align}
Using (\ref{def_M_alpha})-(\ref{def_M_sum}), we have
\begin{align}
\left(\bbk_X^{-1}+\sum_{\ell\in\mathcal{A}^c}\bbh_{\ell,\alpha}^\top(\bbi-\bbd_\ell)\bbh_{\ell,\alpha}\right)^{-1}
&=\Big(\bbm_{\mathcal{A}^c}+\bbm(\alpha)\Big)^{-1} \\
&=\bbm_{\mathcal{A}^c}^{-1/2}\Big(\bbi+\bbm_{\mathcal{A}^c}^{-1/2}\bbm(\alpha)\bbm_{\mathcal{A}^c}^{-1/2}\Big)^{-1}
\bbm_{\mathcal{A}^c}^{-1/2} \label{some_dummy_thing}
\end{align}
Since we have $\lim_{\alpha\rightarrow 0}\bbm(\alpha)=\bzero$,
there exists $\alpha^*$ such that
\begin{align}
\lim_{n \rightarrow
\infty}~\left(\bbm_{\mathcal{A}^c}^{-1/2}\bbm(\alpha)\bbm_{\mathcal{A}^c}^{-1/2}\right)^n=\bzero,
\quad \forall \alpha \in (0,\alpha^*) \label{vanishing}
\end{align}
In view of (\ref{vanishing}), using
Lemma~\ref{lemma_matrix_inverse} in (\ref{some_dummy_thing})
yields
\begin{align}
\Big(\bbm_{\mathcal{A}^c}+\bbm(\alpha)\Big)^{-1}=
\bbm_{\mathcal{A}^c}^{-1/2}\left(\sum_{n=0}^\infty
\left(\bbm_{\mathcal{A}^c}^{-1/2}\bbm(\alpha)\bbm_{\mathcal{A}^c}^{-1/2}\right)^n\right)\bbm_{\mathcal{A}^c}^{-1/2},\quad
\alpha \in(0,\alpha^*)
\end{align}
using which yields
\begin{align}
\lim_{\alpha\rightarrow
0}~\left(\bbk_X^{-1}+\sum_{\ell\in\mathcal{A}^c}\bbh_{\ell,\alpha}^\top(\bbi-\bbd_\ell)\bbh_{\ell,\alpha}\right)^{-1}
&=\lim_{\alpha\rightarrow
0}~\bbm_{\mathcal{A}^c}^{-1/2}\left(\sum_{n=0}^\infty
\left(\bbm_{\mathcal{A}^c}^{-1/2}\bbm(\alpha)\bbm_{\mathcal{A}^c}^{-1/2}\right)^n\right)\bbm_{\mathcal{A}^c}^{-1/2}\\
&=\bbm_{\mathcal{A}^c}^{-1} \label{smthing_implies} \\
&=\left(\bbk_X^{-1}+\sum_{\ell\in\mathcal{A}^c}\bbh_\ell^\top
(\bbi-\bbd_\ell)\bbh_\ell\right)^{-1} \label{def_M_sum_implies}
\end{align}
where (\ref{smthing_implies}) is due to the fact that
$\lim_{\alpha \rightarrow 0}\bbm(\alpha)=\bzero$, and
(\ref{def_M_sum_implies}) is due to (\ref{def_M_sum}). Equation
(\ref{def_M_sum_implies}) is the desired end result in
Lemma~\ref{lemma_limiting_inverse}; completing the proof.

\section{Proofs of Theorem~\ref{theorem_inner} and Theorem~\ref{theorem_inner_general}}
\label{sec:proof_of_inner_bound}

We obtain the inner bound for the rate-distortion region
$\mathcal{R}(\bbd)$ by evaluating the Berger-Tung achievable
scheme with jointly Gaussian auxiliary random vectors. For that
purpose, we consider the most general form of the vector Gaussian
CEO model defined by the observations in (\ref{general_form}). In
other words, we first obtain an inner bound for the most general
form given by (\ref{general_form}), i.e., we prove
Theorem~\ref{theorem_inner_general}, and next, show that
Theorem~\ref{theorem_inner} follows from
Theorem~\ref{theorem_inner_general}. Let $\mathcal{R}^{\rm
BT}(\bbd)$ denote the Berger-Tung inner bound. $\mathcal{R}^{\rm
BT}(\bbd)$ is given by the union of rate tuples $(R_1,\ldots,R_L)$
satisfying~\cite{thesis_tung}
\begin{align}
\sum_{\ell\in\mathcal{A}}R_\ell \geq
I(\bbx;\{U_\ell\}_{\ell\in\mathcal{A}}|\{U_\ell\}_{\ell\in\mathcal{A}^c})
+\sum_{\ell\in\mathcal{A}}I(\bby_\ell;U_\ell|\bbx)
\label{rate_upper_bound}
\end{align}
for all $\mathcal{A}\subseteq\{1,\ldots,L\}$, where the union is
over all $(U_1,\ldots,U_L)$ satisfying the Markov chain
\begin{align}
U_j \rightarrow \bby_j \rightarrow \bbx \rightarrow \bby_k
\rightarrow U_k,\quad j\neq k \label{ach_MC}
\end{align}
and the distortion constraint
\begin{align}
\mmse(\bbx|U_1,\ldots,U_L) \preceq \bbd \label{BT_4}
\end{align}
We select the auxiliary random variables $\{U_\ell\}_{\ell=1}^L$
as follows
\begin{align}
U_\ell=\bby_\ell+\bar{\bbn}_\ell,\quad \ell=1,\ldots,L
\label{auxiliary_rvs}
\end{align}
where $\{\bar{\bbn}_\ell\}_{\ell=1}^{L}$ are zero-mean independent
Gaussian random vectors with covariance matrices
$\{\bar{\bbsigma}_\ell\}_{\ell=1}^L$, and are independent of
$\{\bby_\ell\}_{\ell=1}^L,\bbx$. We assume that the covariance
matrices $\{\bar{\bbsigma}_\ell\}_{\ell=1}^L$ are strictly
positive definite, i.e., we have $\bar{\bbsigma}_\ell\succ
\bzero,~\forall\ell\in\{1,\ldots,L\}$. This assumption arises from
the fact that if one of these matrices is singular, for example,
if $\bar{\bbsigma}_\ell$ is singular, then, as we will show soon,
the corresponding MMSE matrix $\mmse(\bby_\ell|\bbx,U_\ell)$ will
be singular as well, and consequently,
$I(U_\ell;\bby_\ell|\bbx)\rightarrow \infty$. When the auxiliary
random variables $\{U_\ell\}_{\ell=1}^L$ are selected to be
Gaussian as in (\ref{auxiliary_rvs}), the rate bound in
(\ref{rate_upper_bound}) becomes
\begin{align}
\sum_{\ell\in\mathcal{A}}R_\ell \geq
\frac{1}{2}\log\frac{|\mmse(\bbx|\{U_\ell\}_{\ell\in\mathcal{A}^c})|}{|\mmse(\bbx|\{U_\ell\}_{\ell=1}^L)|}
+\sum_{\ell\in\mathcal{A}}\frac{1}{2}
\log\frac{1}{|\mmse(\bby_\ell|\bbx,U_\ell)|}
\label{rate_upper_bound_Gaussian}
\end{align}
where, as it will become clear soon, all MMSE matrices are
strictly positive definite; implying that the rate bounds in
(\ref{rate_upper_bound_Gaussian}) are finite.

Next, we evaluate the MMSE terms in
(\ref{rate_upper_bound_Gaussian}). Using the definition of
auxiliary random variables in (\ref{auxiliary_rvs}), we have (see
(\ref{MMSE_matrix_reference_1}) and
(\ref{MMSE_matrix_reference_2}) in Appendix~\ref{appendix_MMSE})
\begin{align}
\bbd_\ell&\triangleq
\mmse(\bby_\ell|\bbx,U_\ell)\\
&=\left(\bbi+\bar{\bbsigma}_\ell^{-1}\right)^{-1},\quad
\ell=1,\ldots,L \label{individual_MMSE}
\end{align}
where $\bbd_\ell$ satisfies the following orders
\begin{align}
\bzero\prec \bbd_\ell \preceq \bbi \label{orders_of_D_ell}
\end{align}
where the upper bound on $\bbd_\ell$ follows from the following
fact
\begin{align}
\mmse(\bby_\ell|\bbx,U_\ell)&=\mmse(\bbn_\ell|\bbn_\ell+\bar{\bbn}_\ell)\preceq
\bbi
\end{align}
Using (\ref{individual_MMSE}), we have
\begin{align}
\bar{\bbsigma}_\ell=\left(\bbd_\ell^{-1}-\bbi\right)^{-1},\quad
\ell=1,\ldots,L \label{alternative_def}
\end{align}
Next, we evaluate the MMSE matrices
$\mmse(\bbx|\{U_\ell\}_{\ell\in\mathcal{A}^c})$ as follows (see
(\ref{MMSE_matrix_reference_1}) and
(\ref{MMSE_matrix_reference_2}) in Appendix~\ref{appendix_MMSE})
\begin{align}
\mmse(\bbx|\{U_\ell\}_{\ell\in\mathcal{A}^c})&=\left(\bbk_X^{-1}
+\sum_{\ell\in\mathcal{A}^c}\bbh_\ell^\top(\bbi+\bar{\bbsigma}_\ell)^{-1}\bbh_\ell\right)^{-1}\\
&= \left(\bbk_X^{-1}
+\sum_{\ell\in\mathcal{A}^c}\bbh_\ell^\top(\bbi-\bbd_\ell)\bbh_\ell\right)^{-1}\label{dummy_identity_implies}
\end{align}
where we used the following identity
\begin{align}
\left(\bbi+\bar{\bbsigma}_\ell\right)^{-1} =\bbi-\bbd_\ell
\label{dummy_identity}
\end{align}
which can be shown by using (\ref{alternative_def}). Hence, using
(\ref{individual_MMSE}) and (\ref{dummy_identity_implies}) in
(\ref{rate_upper_bound_Gaussian}), we obtain the inner bound as
the union of rate tuples $(R_1,\ldots,R_L)$ satisfying
\begin{align}
\sum_{\ell\in\mathcal{A}}R_\ell \geq
\frac{1}{2}\log\frac{\left|\left(\bbk_X^{-1}+\sum_{\ell\in\mathcal{A}^c}\bbh_\ell^\top
(\bbi-\bbd_\ell)\bbh_\ell\right)^{-1}\right|}
{\left|\left(\bbk_X^{-1}+\sum_{\ell=1}^L \bbh_\ell^\top
(\bbi-\bbd_\ell)\bbh_\ell\right)^{-1}\right|}
+\sum_{\ell\in\mathcal{A}}\frac{1}{2} \log\frac{1}{|\bbd_\ell|}
\label{rate_upper_bound_Gaussian_final}
\end{align}
for all $\mathcal{A}\subseteq \{1,\ldots,L\}$, where the union is
over all positive semi-definite matrices
$\{\bbd_\ell\}_{\ell=1}^L$ satisfying
\begin{align}
\left(\bbk_X^{-1}+\sum_{\ell=1}^L \bbh_\ell^\top
(\bbi-\bbd_\ell)\bbh_\ell\right)^{-1}&\preceq
\bbd \label{dummy_identity_implies_implies}\\
\bzero \preceq \bbd_\ell &\preceq \bbi,\quad \ell=1,\ldots,L
\label{orders_of_D_ell_implies}
\end{align}
where the first constraint in
(\ref{dummy_identity_implies_implies}) is obtained by using
(\ref{dummy_identity_implies}) in (\ref{BT_4}), and the second
constraint in (\ref{orders_of_D_ell_implies}) comes from
(\ref{orders_of_D_ell}). Hence, in view of
(\ref{rate_upper_bound_Gaussian_final})-(\ref{orders_of_D_ell_implies}),
we obtain the inner bound given in
Theorem~\ref{theorem_inner_general}; completing the proof.

Next, we show that Theorem~\ref{theorem_inner} follows from
Theorem~\ref{theorem_inner_general}. We note that the observations
in (\ref{aligned_observations}) are equivalent to the general form
of the observations in (\ref{general_form}), when one sets
$\bbh_\ell=\bbsigma_\ell^{-1/2},~\ell=1,\ldots,L$. Using this
observation in
(\ref{rate_upper_bound_Gaussian_final})-(\ref{orders_of_D_ell_implies})
in conjunction with the definition
$\bbd_\ell=\bbsigma_\ell^{-1/2}\tilde{\bbd}_\ell\bbsigma_\ell^{-1/2},~\ell=1,\ldots,L$,
one can get the inner bound in Theorem~\ref{theorem_inner};
completing the proof.

\bibliographystyle{unsrt}
\bibliography{IEEEabrv,references2}

\begin{thebibliography}{10}

\bibitem{CEO_Introduction}
T.~Berger, Z.~Zhang, and H.~Viswanathan.
\newblock The {CEO} problem.
\newblock {\em {IEEE} Trans. Inf. Theory}, 42(3):887--902, May 1996.

\bibitem{Quadratic_Gaussian_CEO}
H.~Viswanathan and T.~Berger.
\newblock The quadratic {G}aussian {CEO} problem.
\newblock {\em {IEEE} Trans. Inf. Theory}, 43(5):1549--1559, Sep. 1997.

\bibitem{Oohama_CEO}
Y.~Oohama.
\newblock Rate-distortion theory for {G}aussian multiterminal source coding
  systems with several side informations at the decoder.
\newblock {\em {IEEE} Trans. Inf. Theory}, 51(7):2577--2593, Jul. 2005.

\bibitem{Prabhakaran_CEO}
V.~Prabhakaran, D.~Tse, and K.~Ramchandran.
\newblock Rate region of the quadratic {G}aussian {CEO} problem.
\newblock In {\em IEEE Intnl. Symp. Inf. Theory}, page 119, Jun. 2004.

\bibitem{thesis_tung}
S.-Y. Tung.
\newblock {\em Multiterminal source coding}.
\newblock PhD thesis, Cornell University, Ithaca, NY, 1978.

\bibitem{Chen_alternative}
J.~Wang, J.~Chen, and X.~Wu.
\newblock On the sum rate of {G}aussian multiterminal source coding: {N}ew
  proofs and results.
\newblock {\em {IEEE} Trans. Inf. Theory}, 56(8):3946--3960, Aug. 2010.

\bibitem{Shamai_MIMO}
H.~Weingarten, Y.~Steinberg, and S.~Shamai (Shitz).
\newblock The capacity region of the {G}aussian multiple-input multiple-output
  broadcast channel.
\newblock {\em {IEEE} Trans. Inf. Theory}, 52(9):3936--3964, Sep. 2006.

\bibitem{MIMO_BC_Secrecy}
E.~Ekrem and S.~Ulukus.
\newblock The secrecy capacity region of the {G}aussian {MIMO} multi-receiver
  wiretap channel.
\newblock {\em {IEEE} Trans. Inf. Theory}, 57(4):2083--2114, Apr. 2011.

\bibitem{Vector_CEO_sum_rate}
S.~Tavildar and P.~Viswanath.
\newblock On the sum-rate of the vector {G}aussian {CEO} problem.
\newblock In {\em Asilomar Conf. on Signals, Systems and Computers}, pages
  3--7, Oct. 2005.

\bibitem{Chen_Wang_Vector_CEO}
J.~Chen and J.~Wang.
\newblock On the vector {G}aussian {CEO} problem.
\newblock In {\em IEEE ISIT}, pages 2050--2054, Aug. 2011.

\bibitem{Liu_Extremal_Inequality}
T.~Liu and P.~Viswanath.
\newblock An extremal inequality motivated by multiterminal information
  theoretic problems.
\newblock {\em {IEEE} Trans. Inf. Theory}, 53(5):1839--1851, May 2007.

\bibitem{Wagner_Outer_Bound}
A.~B. Wagner and V.~Anantharam.
\newblock An improved outer bound for multiterminal source coding.
\newblock {\em {IEEE} Trans. Inf. Theory}, 54(5):1919--1937, May 2008.

\bibitem{Palomar_Gradient}
D.~P. Palomar and S.~Verdu.
\newblock Gradient of mutual information in linear vector {G}aussian channels.
\newblock {\em {IEEE} Trans. Inf. Theory}, 52(1):141--154, Jan. 2006.

\bibitem{Ekrem_Ulukus_Alternative}
E.~Ekrem and S.~Ulukus.
\newblock An alternative proof for the capacity region of the degraded
  {G}aussian {MIMO} broadcast channel.
\newblock {\em {IEEE} Trans. Inf. Theory}.
\newblock To appear. Also available at [arXiv:1002.4022].

\bibitem{horn_johnson_book1}
R.~A. Horn and C.~R. Johnson.
\newblock {\em Matrix Analysis}.
\newblock Cambridge, 1985.

\bibitem{Dembo}
A.~Dembo.
\newblock Information inequalities and uncertainty principles.
\newblock Tech. Rep., Dept. Statist., Stanford Univ., Stanford, CA., 1990.

\bibitem{Dembo_Cover}
A.~Dembo, T.~M. Cover, and J.~A. Thomas.
\newblock Information theoretic inequalities.
\newblock {\em {IEEE} Trans. Inf. Theory}, 37(6):1501--1518, Nov. 1991.

\bibitem{Liu_Compound}
H.~Weingarten, T.~Liu, , S.~Shamai (Shitz), Y.~Steinberg, and P.~Viswanath.
\newblock The capacity region of the degraded multiple-input multiple-output
  compound broadcast channel.
\newblock {\em {IEEE} Trans. Inf. Theory}, 55(11):5011--5023, Nov. 2009.

\bibitem{Welsh_Matroid}
D.~J.~A. Welsh.
\newblock {\em Matroid Theory}.
\newblock Academic Press, 1976.

\bibitem{Tse_MAC_Fading}
D.~Tse and S.~Hanly.
\newblock Multiple access fading channels-{P}art {I}: {P}olymatroid structure,
  optimal resource allocation and throughput capacities.
\newblock {\em {IEEE} Trans. Inf. Theory}, 44(7):2796--2815, Nov. 1998.

\end{thebibliography}
\end{document}